\documentclass[letterpaper,twocolumn,10pt]{article}
\usepackage{usenix-2020-09}

\usepackage[english]{babel}
\usepackage{graphicx}
\usepackage{ifthen}
\usepackage{tikz}
\usetikzlibrary{positioning, fit, calc, backgrounds}
\usepackage{amsmath}
\usepackage{xspace}
\usepackage[compact]{titlesec}
\usepackage{listings}
\usepackage{multirow}
\usepackage{booktabs}
\usepackage{array}
\usepackage{makecell}
\usepackage{mdframed}
\usepackage{enumitem}
\usepackage[font=small,skip=0pt, textfont=normalfont]{caption}
\usepackage{tightenum}
\usepackage{subcaption}
\usepackage{comment}

\usepackage{soul}

\usepackage{etoolbox}
\makeatletter
\patchcmd{\maketitle}
	{\@maketitle}
	{\vspace{-6em}\@maketitle\vspace{-2em}}%
	{}
	{}
\makeatother

\definecolor{dkgreen}{rgb}{0,0.6,0}
\definecolor{dred}{rgb}{0.545,0,0}
\definecolor{dblue}{rgb}{0,0,0.745}
\definecolor{lgrey}{rgb}{0.9,0.9,0.9}
\definecolor{gray}{rgb}{0.4,0.4,0.4}
\definecolor{darkblue}{rgb}{0.0,0.0,0.6}

\newcommand{\mtp}{MTP\xspace}
\newcommand{\quic}{QUIC-Lite\xspace}
\newcommand{\roce}{RoCEv2\xspace}
\newcommand{\mtpdpdk}{\mtp-DPDK\xspace}
\newcommand{\mtpxdp}{\mtp-XDP\xspace}
\newcommand{\mtpfpga}{\mtp-FPGA\xspace}
\newcommand{\sref}[1]{\S\ref{#1}}
\newcommand{\signpost}[1]{\textbf{#1.}}

\newcommand{\code}[1]{{\small \textsf{#1}}}

\newcommand{\eat}[1]{}
\makeatletter
\newcommand{\pcite}[1]{{\let\cite@adjust\relax\cite{#1}}}
\makeatother

\def\exammode{comment}
\ifthenelse{\equal{\exammode}{nocomment}}
  {%
    \newcommand{\todo}[1]{}
    \newcommand{\stodo}[1]{}
    \newcommand{\htodo}[1]{}
    \newcommand{\temptodo}[1]{}
  }
  {%
    \newcommand{\todo}[1]{{\color{red}[#1]}}
    \newcommand{\stodo}[1]{{\color{red}[#1]}}
    \newcommand{\temptodo}[1]{{\color{red}[#1]}}
    \newcommand{\htodo}[1]{}
  }

\lstdefinelanguage{cpp}{
  basicstyle=\scriptsize \ttfamily \color{black} \bfseries,
  breakatwhitespace=false,
  breaklines=true,
  captionpos=b,
  commentstyle=\color{dkgreen},
  deletekeywords={...},
  frame= none,
  framesep = 2pt,
  framexrightmargin=-4pt,
  language=C++,
  keywordstyle=\color{dblue},
  morekeywords={event, incoming, app_event, instr_t, addr_t, int8, int32, uint16, uint32, uint8, uint64, list, event_t, event_list, fid_t, extract, set_fid, app_req_t, context, timer_t, dispatch, min, max, unseg_data, pkt_event, set_duration, start, mili, add, pkt_bp, checksum16_t, data_t, bit, pkt_t, sliding_wnd, buffer_id_t, data_event, set, first_unset, new_tx_ordered_data, add_tx_data, pkt_gen, seg_rule},
  identifierstyle=\color{black},
  stringstyle=\color{blue},
  numbersep=3pt,                  numberstyle=\scriptsize\color{black}\bfseries,
  xleftmargin=3pt,
  rulecolor=\color{black},
  showspaces=false,
  showstringspaces=false,
  showtabs=false,
  stepnumber=1,
  tabsize=5,
  belowskip=-15pt,
  aboveskip=10pt,
  title=\lstname,
  escapeinside={@@},
}

\lstdefinestyle{mtpfig}{
  language=cpp,
  basicstyle=\fontsize{7pt}{7.4pt}\selectfont\bfseries\ttfamily,
  keywordstyle=\color{dblue}\bfseries,
  commentstyle=\color{dkgreen}\itshape,
  identifierstyle=\color{black},
  morekeywords={net_event, timer_event, pkt_parser, app_parser},
  numbers=none,
  frame=none,
  columns=flexible,
  keepspaces=true,
  showstringspaces=false,
  aboveskip=2pt,
  belowskip=0pt,
  escapeinside={@@},
}
\definecolor{sneedFill}{RGB}{250,224,192}

\newcommand{\sneed}[1]{\tikz[baseline=(SN.base)]\node[fill=sneedFill, rounded corners=1.5pt, inner xsep=2pt, inner ysep=0.7pt] (SN) {\fontsize{7pt}{7.4pt}\selectfont\bfseries\ttfamily #1};}
\newcommand{\lkw}[1]{{\color{dblue}\bfseries #1}}

\lstdefinestyle{cppbox}{
  language=cpp,
  frame=none,
  framesep=2pt,
  framexrightmargin=-4pt,
  xleftmargin=12pt,
  framexleftmargin=12pt,
  xrightmargin=0pt,
  aboveskip=0.3em,
  belowskip=0.2em,
}

\begin{document}

\title{\Large \bf Transport Programs: Specifying The Missing Layer Between Protocols and Stacks}

\author{
  \parbox{\textwidth}{\centering\normalsize\rm
    Pedro~Mizuno$^{1}$, Kimiya~Mohammadtaheri$^{1}$, Linfan~Qian$^{1\dagger}$, Andrew~Qiu$^{1}$, Ashely~Shen$^{1}$, Joshua~Johnson$^{1\dagger}$, Danny~Akbarzadeh$^{1}$, Chris~Neely$^{2}$, Mario~Baldi$^{3}$, Ahmed~Saeed$^{4}$, Nachiket~Kapre$^{1}$, Mina~Tahmasbi~Arashloo$^{1}$\\[0.3em]
    \it $^{1}$University of Waterloo \quad $^{2}$AMD \quad $^{3}$NVIDIA \quad $^{4}$Georgia Tech}
}

\maketitle
\begingroup
\renewcommand{\thefootnote}{\ensuremath{\dagger}}
\footnotetext{Work done while at the University of Waterloo.}
\endgroup

\begin{abstract}
Running code has long served two roles in transport protocol design: a medium for protocol development and a precise behavioral specification. Modern stacks increasingly struggle to serve either role because protocol logic is interwoven with platform-specific machinery for high performance. Changing or comparing protocol behavior therefore requires deep stack expertise, hindering protocol evolution and obscuring interoperability problems.

We propose transport programs, a specification layer between natural-language protocol standards and stack implementations. Written in MTP, a stack-independent language, these programs state protocol decisions precisely while leaving execution machinery to each stack.
MTP provides an explicit interface between protocol and stack experts: transport programs can run on MTP-compatible stacks, describe existing fixed-function implementations for behavioral comparison, and structure tasks for AI agents.
We demonstrate MTP's benefits with three \mtp-compatible stacks, two software and an FPGA prototype, on which we run programs for TCP, Homa, lightweight QUIC, and RoCEv2, comparing MTP descriptions of two TCP stacks to reveal interoperability issues, and creating an \mtp{}-structured agentic pipeline that derives from RFCs a TCP MTP program that interoperates with NS3's TCP.

\end{abstract}

\section{Introduction}
\label{sec:intro}

\begin{quote}
\emph{``We reject: kings, presidents, and voting.
We believe in: rough consensus and running code.''}

\hfill ---David Clark, 1992.
\vspace{-0.08in}
\end{quote}

This credo captures the IETF’s approach to protocol development: specifications, written in natural language, are shaped through rough consensus and refined through implementation and operational experience.
New behavior was proposed and proven as changes to code, which the standards then integrated, and where the standards were ambiguous or deliberately left things open, the code was the precise statement of what an endpoint would do.
For transport protocols, this process remained effective for decades because implementations shared a common shape. That is, a TCP implementation generally looked recognizably like another TCP implementation, with most living in software kernels with broadly similar structures.
So, it was feasible for a protocol change, proven through implementation, to be read, ported, and adopted by the rest, for engineers to identify divergent interpretations and correct them through interoperability testing.

Today, transport protocol implementations can no longer play either role.
With the breakdown of Dennard scaling and the slowdown of Moore's law, modern transport implementations are organized around sustaining the highest packet and data rates the platform can support. So, the structure of the code reflects the platform rather than the protocol, and the protocol logic is interwoven with the machinery that sustains those rates, i.e., memory management, synchronization, batching, pipelining, and placement of work in processing domains where it can be executed most efficiently.
The platforms where transport protocols run have also significantly increased in diversity: the same protocol can have implementations in the kernel, in user space, in eBPF, on SmartNICs, and in fixed-function accelerators, and even a single implementation may span several of these platforms \cite{mtcp,chen2025etran,tonic,linux_kernel_devmem}.

Such implementations can no longer be the main enironment for developing and improving protocols because the cost of changing a protocol inside a modern stack has grown too high.
While modifying and developing protocols through code always required knowing the stack, protocol and stack expertise have since grown apart. Today, the stack knowledge a protocol change requires is deep, different for every target, and unrelated to the decision the protocol expert is making.
For example, even a semantically small protocol change to TCP's loss recovery, such as adding a NACK, requires extending per-flow data structures with optimized linked-list, adding traversal code for the corresponding list, and integrating it into the main busy loop in mTCP \cite{mtcp}; and, in eTran, coordinated changes across the eBPF hooks and user-space threads that split loss recovery, and the maps, dummy packets, and packet metadata through which they coordinate \cite{chen2025etran}.
Essentially, the two kinds of expertise, each now a deep specialty in its own right, should either still reside in the same person or come together in a tightly integrated team that interacts through an informal, error-prone translation between natural-language documents and each individual platform's machinery.

Implementations cannot be the main source of truth about protocol behavior either. While the code is, technically, still a precise statement of what an endpoint would do, protocol logic is scattered and interwoven through the machinery, and extracting it out of the stack takes the same deep, stack-specific knowledge described above.
So catching divergent interpretations across two stacks is increasingly difficult.
In QUIC, for example, many contradictions and conformance errors have escaped years of pairwise interoperability testing, and found only when researchers built a formal model of the English specification by hand \cite{mcmillan-quic}.
In two otherwise conforming implementations, different choices for timers during handshake caused handshake failures under loss \cite{huitema-timeouts}.
As such, with today's implementations as the only precise description of protocol behavior, divergences surface late, on finished implementations, and case by case.
This is costly but patchable in software. For emerging hardware transports, where the protocol is specified by a consortium in natural-language documents and implemented independently in hardware in each vendor’s NICs and stacks \cite{UET}, some divergences would only be found after the design is committed to silicon.

In this paper, we propose a new specification layer between the protocol standards and the stack implementations.%
In particular, we propose that a specification in this layer be a \emph{transport program}: code written in a stack-independent language, which we call \mtp{}, stating a protocol's decisions in full detail without including any of the stack's machinery.

Such a program can play both of the roles described above, and more. First, protocol experts can write and change it in a stack-agnostic manner, and even run it on \mtp{}-compatible stacks without stack expertise.
Second, whether or not a stack is \mtp{}-compatible, the program can precisely describe the stack's protocol behavior, and can be used to compare two stacks for interoperability problems or against another \mtp{} program accompanying a standards document as the precise statement of intended behavior, for compliance problems.
Finally, it can provide structure, including bounded tasks with checkable outputs, for AI agents: whether deriving an \mtp{} program from an RFC, checking a stack's behavior against one, or optimizing its implementation in a particular stack.

Designing such a layer is challenging because a transport program should state only protocol decisions, but protocol decisions are increasingly influenced by a stack's capabilities and constraints. Homa's kernel implementation, for example, adapted its data-packet format so that NIC segmentation offload, built for TCP, could segment its packets \cite{homa-linux}.
So the abstractions have to capture the full diversity of protocol decisions, while still letting a stack state its constraints against them.
Moreover, protocol decisions can influence stack decisions. For example, how a stack buffers received data could change based on whether the protocol delivers fixed-size messages or a byte stream.
So the abstractions should capture protocol decisions in a form that stacks can use to make their own decisions well.

Addressing these challenges comes down to carefully deciding, for each part of transport, what should be stated in the program and what should be left to the stack.
We organize these decisions around two observations.
First, while transports are diverse in their decisions, they are uniform in structure. They follow stateful event processing, where upper-layer requests, packet arrivals, and timer expirations update the protocol's state and trigger actions such as sending packets, reassembling and delivering data, and adjusting timers.
So capturing the diversity of protocol decisions breaks into four design problems: abstractions for events, state, processing logic, and actions.

Second, the events, the state, and, in particular, the actions, which in existing stacks are integrated with heavy machinery, can be abstracted in a declarative way, i.e., stated through their properties and parameters rather than implemented. This helps state protocol decisions while keeping the machinery out of the program. Each stack can realize a program's declarations with its own machinery based on the declared properties, and state its constraints against them.
What remains is a simple imperative processing logic that states the protocol's run-time decisions, mapping events and state to updated state and actions. Specifically, with the actions declared rather than implemented in the program, the logic is simple enough for protocol experts to navigate, and for a compiler to extract the remaining properties a stack may need, e.g., which events read and write which state, or what triggers each packet transmission, with standard static analysis.

We demonstrate the benefits of \mtp{} as a language-based specification layer between standards and stacks in three parts:
\begin{itemize}[leftmargin=*,itemsep=1pt,topsep=2pt,parsep=0pt]
\item \emph{Separation of concerns and reuse (\sref{sec:benefit1})}.
We realize this benefit by building \mtp{}-compatible stacks, where a compiler can take any transport program within the stack's stated constraints and generate code that builds into the stack, with no changes to the stack's own code. We have built two such software stacks, \mtpdpdk{} and \mtpxdp{}, built by refactoring mTCP and eTran, and an FPGA prototype of a hardware one, \mtpfpga{}.
We show that transport programs for TCP \cite{tcp-rfc}, Homa \cite{homa}, and a lightweight QUIC \cite{quic-paper}, can compile to the software stacks, including protocols mTCP and eTran never implemented, at performance comparable to, and sometimes better than, the hand-built originals, and TCP and RoCE programs run on \mtpfpga{}.

\item \emph{Precise communication of intended behavior (\sref{sec:benefit2})}. We write \mtp{} programs for the TCP implementations of mTCP and eTran and, by analyzing the two programs alone, find real interoperability divergences between the stacks, including a window-scaling mismatch and divergence in handling data and ack packets that cause performance degradation.

\item \emph{Structure for agentic development (\sref{sec:benefit3})}. We create an agentic pipeline that derives a TCP program from RFC 9293 and the documents it references, about a hundred pages, generating an implementation of TCP that interoperates with NS3's native TCP.
\end{itemize}

A specification layer for transport cuts across an entire ecosystem built over decades. A ``full'' evaluation means building every kind of stack, modeling every deployed protocol, and reproducing its accumulated interoperability practice, each of which is a major undertaking of its own.
Abstractions of this kind also settle over years, through the community using and refining them: P4's details have changed considerably over a decade of use, while many of its main ideas such as programmable parsing and match-action processing have stayed \cite{bosshart2014p4,p4}.
We have evaluated \mtp{} as far as one paper can through building and evaluating representative prototype stacks and protocols, and doing interoperability and agentic flow case studies, and believe our results provide compelling evidence that creating such a specification layer is feasible and valuable, with structures and ideas the community can build on and refine.

\begin{figure*}[t]
\centering
\setlength{\abovecaptionskip}{4pt}
\setlength{\belowcaptionskip}{-12pt}
\includegraphics[width=0.9\textwidth]{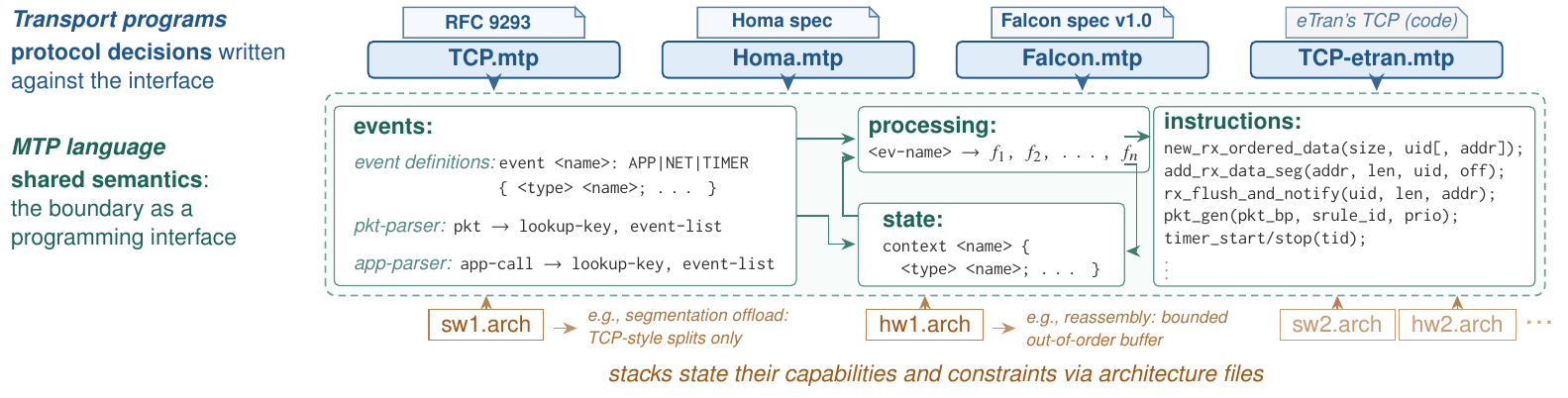}
\caption{Transport programs written in \mtp{} (\sref{sec:mtp}), the specification layer motivated by the
growing divergence between protocol and stack expertise (\sref{sec:motivation}).
Protocol experts state protocol decisions as programs against \mtp{}'s
abstractions, events, state, processing, and transport instructions (top and
center). Stack experts state capabilities and constraints against the same
abstractions, in architecture files (bottom).
Transport programs can program and change the protocol behavior of MTP-compatible stacks or specify the behavior of fixed-function stacks.}
\label{fig:mtp-template}
\end{figure*}

\section{Motivation}
\label{sec:motivation}

As discussed in \sref{sec:intro}, the need for a specification layer between standards and stacks stems from the growing divergence between protocol and stack expertise. We show this divergence more concretely through the decisions each expertise makes (\sref{sec:expertise}). We then discuss the benefits of a language-based specification layer between the two (\sref{sec:why-boundary}) and why no existing abstraction provides one (\sref{sec:statusquo}).

\subsection{The two kinds of expertise}
\label{sec:expertise}

\signpost{Growing depth of protocol expertise} Protocol decisions involve (1) what events the protocol reacts to, i.e., what operations the upper layer (e.g., an application or an RPC library) can invoke, what kinds of packets are allowed, and what timers to maintain, (2) what state to maintain, and at what granularity, and (3) how each event updates that state, what packets to generate in response with what header, payload, rate, and priority, and in what order to reassemble incoming payloads and when to deliver them.

These decisions have grown richer over decades of protocol evolution, and the protocol skills and expertise needed to make them well have deepened accordingly: TCP's loss recovery alone grew from the retransmission timeouts and duplicate acknowledgments of RFC 793 and 5681, through SACK and the scoreboard rules of RFC 6675, to the time-based detection and probes of RACK-TLP (RFC 8985), with congestion control and security rules evolving alongside. And TCP is just one of many: QUIC, Homa, RoCE, SRD, Falcon, and UET each make their own versions of these decisions \cite{quic-paper,homa,roce,srd,falcon,UET}.

\signpost{Growing depth of stack expertise} Stack decisions have grown just as rich, in a different direction. Whether in the kernel, in user-space frameworks like mTCP, or in NIC hardware, they involve (1) how to partition work and state across cores, pipeline stages, and processing domains, (2) how to organize transport data and state in memory, and how cores or pipeline stages share it with minimal synchronization, (3) the per-event execution model, i.e., whether each event is processed to completion or in stages, (4) how to amortize per-packet and per-operation overheads, from batched I/O and coalescing to segmentation offload, and (5) how packet I/O and event notification are implemented, e.g., interrupts versus polling, and how timers are tracked and fired. These decisions are made by reasoning about the platform, e.g., cycles, cache lines, cores, and memory bandwidth, and the workload, e.g., packet rates, flow counts, and latency budgets.
They are also made differently on every platform: the same protocol may run in per-core busy-polling threads with batched I/O \cite{mtcp}, split across eBPF hooks, maps, and user-space threads \cite{chen2025etran}, and as hardware pipelines \cite{tonic}, each organization the product of years of platform-specific engineering.

\subsection{Benefits of language-based specifications}
\label{sec:why-boundary}

A language-based specification layer between standards and stacks provides three main benefits.

\signpost{B1: Separation of concerns and reuse} A stack can implement its side of the boundary once, so that any transport program could run on it. This way, protocol experts can change an endpoint's reactions without understanding the stack's machinery, stack experts can reorganize that machinery without changing protocol behavior, a protocol description can be reused across stacks, and the cost of building and optimizing a stack is amortized across protocols.

\signpost{B2: Accurate communication of intended behavior for interoperability, testing, and verification} A transport program records protocol decisions precisely and free of any stack's machinery. Such programs can accompany the RFC as its reference, or accompany a stack recording its protocol decisions. This enables testing, and potentially verifying a protocol design before any implementation exists, and comparing the protocol behaviors of two stacks for potential divergence.

\signpost{B3: Structure for agentic development} A transport program is written in a well-defined vocabulary of events, states, and instructions, so work on it splits into bounded tasks with mechanically checkable results that agents do their best work on. With a domain-appropriate structure in place, agents can derive a program from an RFC or from an existing stack's code, optimize a program's implementation on a stack, or check a stack's behavior against one.

\subsection{Where existing abstractions fall short}
\label{sec:statusquo}

No existing interface or abstraction lets a program state every decision that is the protocol's to make, while leaving out the stack's decisions and machinery. Below, we describe the closest existing efforts.

\signpost{Partial abstractions} CCP \cite{narayan2018restructuring} and Falcon \cite[Table 3]{falcon} allow writing programs that make (congestion) control decisions without deep knowledge of the stack's transport data path. These works provide evidence that a layer between protocol and stack is valuable. But they carve out only (congestion) control decisions. The rest of the protocol's many decisions stay interwoven with stack machinery. We revisit partial and functional carve-outs in \sref{sec:mtp}.

\signpost{Interfaces within a particular modular stack} Another line of work builds protocol-independent interfaces inside one stack, so that several protocols can share that stack's machinery: Linux, like FreeBSD, keeps packet demultiplexing, receive queueing, timers, and pacing behind internal interfaces that TCP, UDP, and Homa's kernel module can use; FD.io's TLDK packages TCP and UDP as libraries sharing its DPDK-based packet I/O \cite{tldk}; eTran restructures kernel transport around eBPF hooks and user-space threads, shared by its TCP and Homa \cite{chen2025etran}.

These interfaces lower the cost of hosting and changing protocols on one stack, and some of these stacks can even be ported to a new environment wholesale, e.g., the kernel stack running in user space as a library \cite{lkl,fstack}. But even with these interfaces, running code can play neither of the two roles we need (\sref{sec:intro}). Changing the protocol, or extracting all of its decisions, still takes a stack expert because the logic is plugged into these interfaces in bits and pieces, and is still written against that platform's machinery such as buffers, locks, and execution contexts. Even with Linux's interfaces, adding Homa to the kernel took about two years of iteration \cite{homa-linux}. Moreover, the protocol code cannot be reused on another stack without rewriting.

\signpost{Domain-specific languages} Prolac is a language for writing TCP implementations in a readable and extensible form \cite{prolac}. However, Prolac programs also make stack decisions and rely on target-specific C code for system interactions such as buffers and I/O, so they are still tied to the stack they were written for. P4 is the closest example of what we are after, for switches rather than transports: a P4 program precisely describes a switch's packet processing, independent of the switch that runs it \cite{bosshart2014p4}. P4's vocabulary, however, is limited to per-packet operations on headers and limited state, and cannot express the decisions of a transport protocol expert: how application requests, packet arrivals, and timer expirations update rich state, and what packets and data it produces.

\section{Transport Programs}
\label{sec:mtp}
\begin{figure*}[t]
\begin{minipage}[t]{0.46\textwidth}
\begin{lstlisting}[style=mtpfig]
// Homa.mtp (excerpts)
event grant_rcvd : @\sneed{\lkw{net\_event}}@ {
  uint64 rpcid;  uint32 granted_off; }
@\cgap@
event snd_resp : @\sneed{\lkw{app\_event}}@ { uint64 rpcid;
  data_t resp;   /* to send: (address, size)*/ ...}
@\cgap@
event rsnd_fired : @\sneed{\lkw{timer\_event}}@ { }
@\cgap@
context rpc_ctx : @\sneed{\lkw{key\_size}(112b)}@ {
  // key: rpcid + IP address + port, see parsers
  uint32 granted, rcvd;
  @\lkw{timer}\ rsnd\_tmo $\to$ rsnd\_fired;@ ...}
@\cgap@
context grant_ctx : @\sneed{\lkw{global}}@ {
  list<rpc_info> active_rpcs; uint64 total_incoming; ...}
@\cgap@
// packet blueprints (two of several shown)
@\sneed{\lkw{pkt\_bp}\ GrantBP \{}@ ..., uint64 rpcid;  uint32 off; ...}
@\sneed{\lkw{pkt\_bp}\ DataBP \{}@ /*hdr fields*/; @\sneed{data\_t payload;}@ }
\end{lstlisting}
\end{minipage}
\hfill
\begin{minipage}[t]{0.50\textwidth}
\begin{lstlisting}[style=mtpfig]
list<event_t> @\sneed{\lkw{pkt\_parser}\ homa\_pkts}@(pkt p) {
  ...; GrantBP bp;  extract(p, bp); // P4-like syntax
  grant_rcvd ev(bp.rpcid, bp.granted_off);
  @\sneed{set\_ctx\_lookup\_info(ev, rpc\_ctx,}@
    @\sneed{\{bp.rpcid, p.ip.saddr, bp.sport\});}@ ... }
list<event_t> @\sneed{\lkw{app\_parser}\ homa\_app\_calls}@(...) {...}
@\cgap@
// seg rule, used by pkt_gen below
@\sneed{\lkw{seg\_rule}\ r1(x)}@[DataBP::off,x,prev.off+prev.payload.len];
@\cgap@
// processing for grant_rcvd
@\sneed{\lkw{dispatch}}@ homa_dispatch {
  grant_rcvd -> {grant_ep}; ... }
@\cgap@
list<instr_t> grant_ep(grant_rcvd e, rpc_ctx ctx) {
  list<instr_t> out;
  if (e.granted_off > ctx.granted) {
    ctx.granted = e.granted_off;
    DataBP bp;  ... // fill out bp headers
    @\sneed{out.add(pkt\_gen(bp, r1(ctx.snd\_nxt), ...));}@
  ...; return out; }
\end{lstlisting}
\end{minipage}
\vspace{-20pt}
\setlength{\abovecaptionskip}{3pt}
\caption{Code snippets from the MTP program for the Homa protocol, with some of MTP’s core abstractions for events, application data, packet payloads, and transport instructions highlighted in yellow (\sref{sec:event-processing}, \sref{sec:instructions}).}
\label{fig:homa-events}
\vspace{-0.2in}
\end{figure*}
We propose expressing a protocol's behavior as a \emph{transport program} in a domain-specific language, \mtp, that provides an explicit boundary between protocol logic and stack machinery.
As shown in Figure~\ref{fig:mtp-template}, \mtp{} serves \emph{both as a programming interface and as a specification language}.
As a programming interface, it allows a transport program to change the protocol behavior of an \mtp{}-compatible stack within the stack's stated constraints: the protocol expert states how the endpoint should react to application requests, packets, and timers, what state it should keep, and what packets and data it should produce.
The stack expert supplies an architecture file, stating the stack's capabilities and constraints against the \mtp{} abstractions, and the machinery that executes programs.
As a specification language, a program describes an existing stack's behavior in machine-readable form, even when that stack is ``fixed-function''. That is, the stack expert supplies only the architecture file and no execution support. Such programs can be compared, tested, and analyzed independently of stack code, or accompany an RFC as the precise statement of intended behavior.

\signpost{The challenge} To that end, a transport program should satisfy two challenging requirements:

\emph{R1. Express \ul{all} protocol decisions}
A program should be able to specify the endpoint's behavior completely, including choices that standards leave open and adaptations made for particular stacks or workloads, capturing the full diversity of protocol decisions.

\emph{R2. Express \ul{only} protocol decisions}
A program must leave execution mechanisms to the stack. It must still expose what stack experts need: the information to choose and optimize those mechanisms, and the abstractions to state their constraints against so the protocol expert can account for them without understanding the machinery.

\signpost{Choosing the decomposition} Designing a language for such transport programs comes down to carefully deciding, for each part of transport, what should be stated in the program and what should be left to the stack.
One approach is to organize the boundary around functional components, such as congestion control, retransmission, and flow control.
However, each such component interacts with the rest of the transport, e.g., through the packets and application requests that trigger protocol processing, state, and its impacts on actions such as packet generation and timers. Congestion control, for example, reads acknowledgments and ECN marks from incoming packets, keeps its window in state, backs off retransmission timers, and paces outgoing packets. So an abstraction for one component ends up making assumptions that can restrict both other protocol choices or stack implementations. For example, the kernel's pluggable congestion control assumes the TCP Linux stack by construction, and CCP \cite{narayan2018restructuring} assumes a sender-driven protocol with particular congestion signals.

\mtp{} instead builds a general logical abstraction on stateful, event-driven programming. We take transport to be any protocol that lets an upper layer, e.g., an application or an RPC library, exchange arbitrary-sized data over an unreliable packet network. The behavior of such protocols follows a common structural pattern: an input triggers an event, the protocol consults its state, and processing determines state updates and actions. This structural decomposition leaves protocol experts free to organize their programs functionally, with interactions among features expressed within the program. Functional interfaces can still be built, as libraries or abstractions over transport programs.

\signpost{Deriving the boundary}
For each component in the above logical model, i.e., events, state, processing, and actions, we ask two sets of questions, corresponding to the two requirements above. For R1, we ask: what abstractions are needed in a transport program to describe the protocol's decisions in full, and in a form the stack can use to make its own decisions well? For R2, we ask: what stack and workload factors do those decisions depend on, to be stated outside the program?
These questions guide the abstractions developed in the following
sections, summarized in Figure~\ref{fig:mtp-template}:

\emph{Events}(\sref{sec:events}) represent occurrences to which the protocol reacts: upper-layer requests, incoming packets, and timer expirations. They including the metadata needed to react to the event. The program declares event types and specifies how concrete inputs map to them. How inputs are received and processing is invoked is the stack's decision.

\emph{State} (\sref{sec:state}) is the information kept across events that determines the protocol's response. \mtp{} groups state variables into \emph{contexts} at protocol-defined granularities, such as a connection, an RPC, or the endpoint as a whole. The program defines the variables and which instances are accessed by each event. How instances are stored, located, and kept consistent is the stack's decision.

\emph{Processing} (\sref{sec:processing}) connects the event and its state to the protocol's response. A program maps each event to a chain of functions that read event metadata and contexts, update state, and determine actions. The program specifies these computations precisely. How they execute, across threads, cores, or pipeline stages, is the stack's decision.

\emph{Transport instructions} (\sref{sec:instructions}).
Processing requests actions through transport instructions, summarized in Table~\ref{tab:instructions}. Each specifies an operation and its protocol-defined parameters, such as generating packets, reassembling or delivering data, or starting a timer.
Designing transport instructions is particularly challenging as they must precisely express the protocol's intended action without incorporating the stack machinery that performs it. This requires more than saying which action to take. For example, a packet-generation instruction must specify how to construct the header and payload as well as segmentation behavior while preserving opportunities for I/O optimizations. Or, a reassembly instruction must specify data placement and delivery semantics without fixing a buffer representation or I/O mechanisms. Each action therefore requires its own abstraction, with parameters that capture protocol decisions and constraints that describe what the stack supports.

\section{Stateful Transport Event Processing}
\label{sec:event-processing}

\mtp{} describes an endpoint's protocol behavior through event and state declarations, parsing and state lookup information, and processing functions. These expose the protocol-level computation and state dependencies that the stack must preserve while leaving its execution strategy open. This section describes \mtp{}'s abstractions for events, state, and processing using a Homa grant as a running example.

\subsection{Events and Input Mapping}
\label{sec:events}

The protocol expert determines which inputs trigger a reaction and what information is needed to determine that reaction. \mtp{} expresses these choices through event declarations, each specifying a name, metadata fields, and a source: an application operation (\code{app\_event}), a network packet (\code{net\_event}), or a timer expiration (\code{timer\_event}). A Homa grant event, for example, includes an RPC identifier and a granted offset (Figure~\ref{fig:homa-events}).

Packet parsers, specified using P4-like syntax \cite{bosshart2014p4}, and application parsers specify how concrete inputs produce these events and populate their metadata. Separating event declarations from input mappings allows different packet encodings or application calls to produce the same event when they require the same reaction (\sref{app:event-mapping}). For events involving payloads, \mtp{} provides \code{data\_t}, a data reference containing an abstract address (\code{addr\_t}) and a length. This identifies the relevant bytes without prescribing their physical location or layout.

The event declarations and parsers provide the information the stack experts need to make decisions about how inputs are received, parsed and presented for processing. The declared event source separates inputs that may take different receiving machinery, and the parsers specify the information that the machinery should extract.
In the architecture file, the stack experts can state the supported application interfaces and parsing capabilities, including limits on header length and field extraction, for the protocol expert to take into account in their decisions (\sref{app:event-mapping}).

\subsection{State and Context Lookup}
\label{sec:state}

The protocol expert determines which information persists across events, at what granularity it is maintained, and which events access it. Homa, for example, maintains per-RPC state for individual transfers and global state for grant scheduling.

A \emph{context declaration} specifies a group of typed state variables
and the size of the key used to identify an instance. Each instance
holds a separate set of values, such as the state of one RPC. The
program also specifies which instances are accessed by each event: the packet
or application parser uses \code{set\_ctx\_lookup\_info} to identify
the context and expresses which packet fields or call arguments form the lookup key.
Key construction belongs in the parser because different inputs may
identify the same instance through different fields.
For a Homa grant, the key identifies the corresponding RPC.

The stack expert chooses how to store context instances and keep them
consistent during concurrent processing.
Context declarations and parsers establish the logical organization and sharing of state.
This information lets the stack expert organize processing to avoid unsafe concurrent accesses as efficiently as possible.
For example, it can partition Homa's per-RPC context instances across cores and use
the lookup keys to steer each RPC's events to its core and avoid synchronization.
Homa's global grant-scheduling state, however, may be accessed by events running on
several cores, requiring synchronization such as locks around
conflicting accesses. The processing functions further identify which
fields each event reads and writes, allowing the stack to choose the
appropriate synchronization for its execution model
(\sref{sec:processing}, \sref{sec:targets}).
Additional protocol and workload factors informing these implementation choices are
detailed in \sref{app:state-factors}

In the architecture file, the stack expert can state the memory resources
available for protocol state, including their capacity and kinds.
These constraints bound how much state a program
can maintain, for the protocol expert to take into account in
choosing how they organize the protocol state.

\begin{table}[t]
\centering
\footnotesize
\setlength{\belowcaptionskip}{-15pt}
\setlength{\abovecaptionskip}{2pt}
\begin{tabular}{@{}ll@{}}
\toprule
\multirow{3}{*}{\makecell[l]{\textbf{Data Reassembly}\\
    \textbf{(\sref{sec:data-instrs})}}}
 & new\_rx\_ordered\_data(size, uid[, addr]) \\
 & add\_rx\_data\_seg(addr, len, uid, off) \\
 & rx\_flush\_and\_notify(uid, len, addr) \\
\midrule
\multirow{4}{*}{\makecell[l]{\textbf{Packet Generation}\\
    \textbf{(\sref{sec:pktgen-instrs})}}}
 & pkt\_gen(pkt\_bp, srule\_id[, prio, qid]) \\
 & new\_tx\_ordered\_data(size, uid[, addr]) \\
 & add\_tx\_data(addr, len, uid) \\
 & tx\_flush\_and\_notify(uid, len) \\
\midrule
\multirow{2}{*}{\makecell[l]{\textbf{Timers \& Context}\\
    \textbf{Management (\sref{sec:timer-instrs})}}}
 & timer\_start/stop(tid) \\
 & new\_ctx(init\_vals), del\_ctx() \\
\midrule
\multirow{2}{*}{\makecell[l]{\textbf{Packet Scheduling}\\
    \textbf{(\sref{sec:timer-instrs})}}}
 & set\_queue\_rate(qid, rate) \\
 & set\_queue\_prio(qid, prio) \\
\midrule
\makecell[l]{\textbf{Other} \textbf{(\sref{app:other_instrs})}}
 & notify(msg) \\
\end{tabular}
\caption{\mtp{} transport instructions (\sref{sec:instructions}).
They abstract actions with parameters needed for the program to express the action outcome, leaving its implementation to the stack, which can use those
parameters to make their decisions and state their constraints.}
\label{tab:instructions}
\end{table}

\subsection{Processing Logic}
\label{sec:processing}

The protocol expert determines how an event and the relevant state
produce state updates and transport actions. We discuss the processing logic
making this determination here, and transport actions in \sref{sec:instructions}. An \mtp{} program
includes a dispatch table, specifing which functions process each
event type and the order in which they run. These functions receive
the event and its relevant context instances, read and update context
variables, and issue transport instructions.
Function bodies use
C-like imperative code with bounded
loops, similar to other DSLs~\cite{bosshart2014p4,sivaraman2016packet,tonic}.
In the Homa example (Figure~\ref{fig:homa-events}), the dispatch table selects \code{grant\_ep}
for a received grant. The function compares the received offset with
the offset stored in the RPC context. If the grant permits additional
data, it updates that context and issues a packet-generation
instruction.

The stack expert determines, based on the stack's own placement and synchronization strategies,
how this computation executes: where functions run, whether an event is processed to completion in one
execution context or across several stages, and how accesses to
shared state are synchronized.
The program provides the information the stack needs make these decisions.
Specifically, the dispatch table identifies the functions required for each event, and the
chain of function bodies specify the computations and context-field accesses, from which a
compiler can determine read/write dependencies, and use together with the
context lookup rules (\sref{sec:state}) for state placement and synchronization decision.

In the running Homa grant example, with the information in the program, a software multicore stack knows which contexts can be partitioned across cores (the per-RPC context) and can steer each grant to the core holding its RPC context to run \code{grant\_ep} there, in a run-to-completion thread in a kernel-bypass stack or a softirq in the Linux kernel. It also knows, from the issued instruction, that the resulting action is a packet generation, so nothing in the reaction requires another execution context and the chain can run to completion on that core. Whether the packet is then generated inline or by separate machinery, e.g., a pacer thread, remains the stack's decision. \sref{sec:targets} discusses software and hardware stacks with different execution models in more detail.

The architecture file states restrictions on the computation that
events can trigger and the state accessible to that computation, which
the protocol expert must take into account to express the reactions within those limits.
We discuss code optimizations, semantically equivalent rewrites of the processing logic tuned to a stack, in \sref{app:code-opts}.

\section{Transport Instructions}
\label{sec:instructions}

The transport instruction set specifies the actions a protocol wants to be taken
in response to an event: reassembling and delivering received data to the application, generating and
scheduling packets toward the network, and setting the protocol’s
timers and managing its context lifetimes (Table~\ref{tab:instructions}).
The instruction set is fixed and designed to be general: protocol features such as congestion control and loss recovery are implemented by composing these actions in the processing logic for relevant events, without adding feature-specific instructions. Our case studies demonstrate the expressiveness of this instruction set, including a full TCP implementation in \sref{sec:benefit3}.
Loss recovery, for example, determines which bytes to retransmit and when, then
requests their transmission through the same packet-generation instruction used
for transmitting new data in response to application requests.
Our design also allows the instruction set to evolve to meet future protocol and stack requirements.

Each kind of action requires an abstraction that specifies its
outcome while leaving its implementation to the stack.
In this section, we identify, for each action, the parameters needed to express
the protocol's decisions and  how stacks can use those parameters to make their decisions
or state their constraints.

\subsection{Data Reassembly and Delivery}
\label{sec:data-instrs}

The protocol expert determines which received bytes belong to
the same stream or message, where each segment belongs within it,
and when those bytes should become available to the application.
TCP, for example, reassembles an unbounded byte stream per
connection, whereas Homa or RoCEv2 reassembles a bounded message per RPC or WQE.
The reassembly instructions must express these choices without
prescribing how the stack buffers the bytes.

\mtp{} represents each stream or message as a \emph{logical unit
of ordered data}, identified by \code{uid}, with byte positions
specified as offsets from its beginning. The program declares a
unit using \code{new\_rx\_ordered\_data(size, uid[, addr])}.
Its size is either a known length or \code{INF} for an unbounded
stream. The optional address specifies an application-provided
destination known before the data arrives.
For an arriving segment, \code{add\_rx\_data\_seg(addr, len, uid, off)} specifies that the
\code{len} bytes referenced by \code{addr} belong at offset
\code{off} in unit \code{uid}. The reference comes from the
event's data metadata (\sref{sec:events}).
\code{rx\_flush\_and\_notify(uid, len, addr)} specifies that the
next \code{len} bytes following the already delivered data from the unit
should become available to the application at the supplied
destination. Together, these instructions specify the unit's
size, the position of each segment within it, and the progression of
delivery to the application.

The stack expert chooses how to allocate buffer space, track pending
segments, and place bytes in application memory. The instruction
parameters inform these choices directly. For example, a bounded \code{size}
allows the stack to consider allocating space for the whole
message, or and an upfront destination address when declaring the unit signals opportunities
for direct placement.
The stack may instead buffer segments and copy them when delivery is requested.
These choices must preserve the byte positions and delivery
specified by the program.

In the architecture file, the stack expert can state the available buffering memory and
limits on out-of-order reassembly, which the protocol expert should take into account
when deciding receive and recovery behavior. These limits are
expressed over the unit's abstract state $(f,R)$: $f$ is the
number of bytes already made available to the application,
and $R$ is the set of pending byte ranges. Adding a segment
extends $R$, and flushing advances $f$ and removes the delivered
bytes from $R$ (\sref{app:data-factors}).
A stack can bound the number of ranges in $R$ or their maximum
distance beyond $f$, independently of how it stores those
ranges.

\subsection{Packet Generation}
\label{sec:pktgen-instrs}

The protocol expert determines the packets to transmit, i.e., their
header fields and payload bytes. The \code{pkt\_gen} instruction
expresses these decisions through a packet \emph{blueprint} and a
segmentation rule, which together, describe the requested packets without constructing
concrete packet buffers.

A blueprint specifies the order and values of header fields.
It identifies payload bytes through logical data transmission units,
the send-side counterpart of the reassembly units above.
\code{new\_tx\_ordered\_data} declares a unit,
\code{add\_tx\_data} adds application data to it, and
\code{tx\_flush\_and\_notify} retires a prefix that is no longer needed
by the protocol, for example after an acknowledgment. A blueprint
determines payload bytes using \code{data(uid, off, s)}: \code{s}
bytes starting at offset \code{off} in unit \code{uid}.
This reference identifies the bytes without prescribing where
the stack stores them. Blueprints can also omit payloads for
control packets or be nested for more complex payloads, e.g., in QUIC~\cite{quic-paper}.

One \code{pkt\_gen} instruction can request transmission of more data than what fits
in a packet using a  \emph{segmentation rule}, which specifies
how header fields change across the resulting
packets. A rule has the form \code{[field, first, mid, last]},
whose expressions compute the field's value for the first,
intermediate, and final packets. The expressions can use the
blueprint and previously generated packet. A TCP segmentation rule, for
example, advances the sequence number by the preceding packet's
payload length (see \sref{app:seg-rules} for concrete rule examples).

The stack expert can use this information to choose how to buffer pending transmission data,
construct headers, access payload bytes, and submit packets
to the NIC.
It can choose to generate packets in software or invoke a segmentation offload that supports the
specified rules. Representing several packets in one
instruction also gives the stack an opportunity to amortize
header construction, packet allocation, and I/O. Whether and
how to exploit these opportunities remains a stack
implementation decision.
Moreover, in the architecture file, the stack expert can state constraints on the packet formats
the stack can construct or segmentation offloads it supports, which the protocol expert should take into account when deciding header formats and segmentation strategies.

\subsection{Timers, Packet Scheduling, and More}
\label{sec:timer-instrs}

\signpost{Timers} Protocol experts determine which timers the protocol keeps by declaring each timer in a context and associating it with the timer event created on its (\sref{sec:events}), and how they are managed using \code{timer\_start/stop(tid)} instructions.
The stack expert chooses how to track timers and manage their expiry.
the architecture file states the timers the stack supports and their granularity, which the protocol expert takes into account, e.g., in choosing timeout values.

\signpost{Context management} The protocol expert determines when each instance of the protocol's state is created and retired using \code{new\_ctx(init\_vals)} and \code{del\_ctx()} instructions, e.g., at connection setup and teardown. The stack expert chooses how context memory is allocated and reclaimed.

\signpost{Packet scheduling} The protocol expert determines the order and pace of transmissions, e.g., rate limits in some protocols, e.g., BBR and DCQCN, priorities among packets in others, e.g., Homa and NDP. The stack states its available scheduling blocks, queues and available scheduling policies, in its architecture file. The program instantiates them once with \code{register\_pkt\_sched}, directs packets to queues through \code{pkt\_gen}, and adjusts scheduling policies, if available by the stack, through \code{set\_queue\_rate(qid, rate)} and \code{set\_queue\_prio(qid, prio)} at run time. How the queues and policies are realized remains the stack's decision (\sref{app:other_instrs}).

\section{Benefit 1: Separation of concerns \& reuse}
\label{sec:benefit1}
\begin{figure*}
\centering
\setlength{\abovecaptionskip}{5pt}
\setlength{\belowcaptionskip}{-13pt}

\noindent
\begin{minipage}[b]{0.28\textwidth}
  \centering
  \begin{subfigure}[b]{\linewidth}
    \centering
    \includegraphics[width=0.95\linewidth,height=0.164\textheight,keepaspectratio]{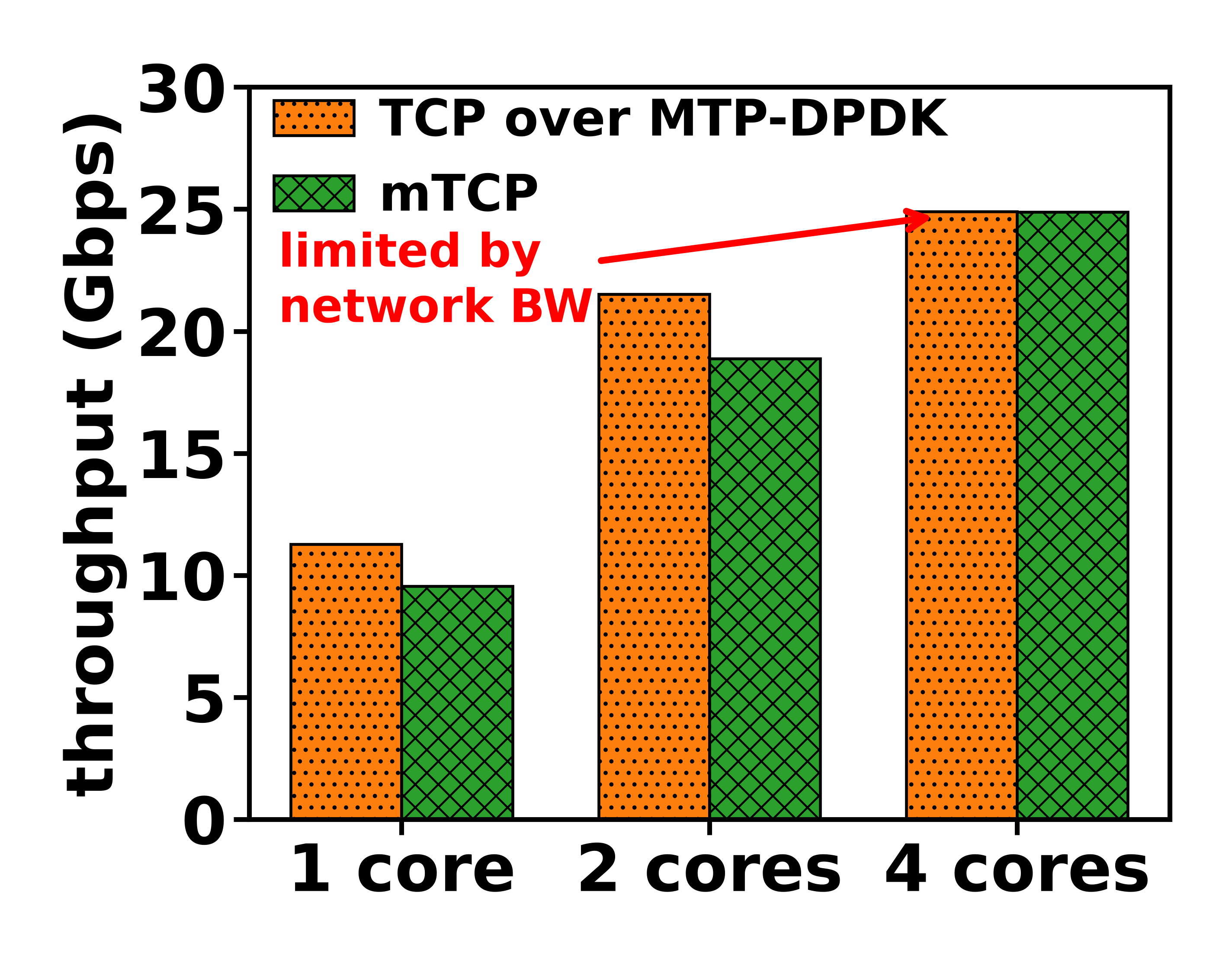}
    \caption{\small \mtpdpdk vs.\ mTCP throughput}
    \label{fig:tcp-dpdk-throughput}
  \end{subfigure}
\end{minipage}%
\hfill%
\begin{minipage}[b]{0.30\textwidth}
  \centering
  \begin{subfigure}[b]{\linewidth}
    \centering
    \includegraphics[width=\linewidth,height=0.164\textheight,keepaspectratio]{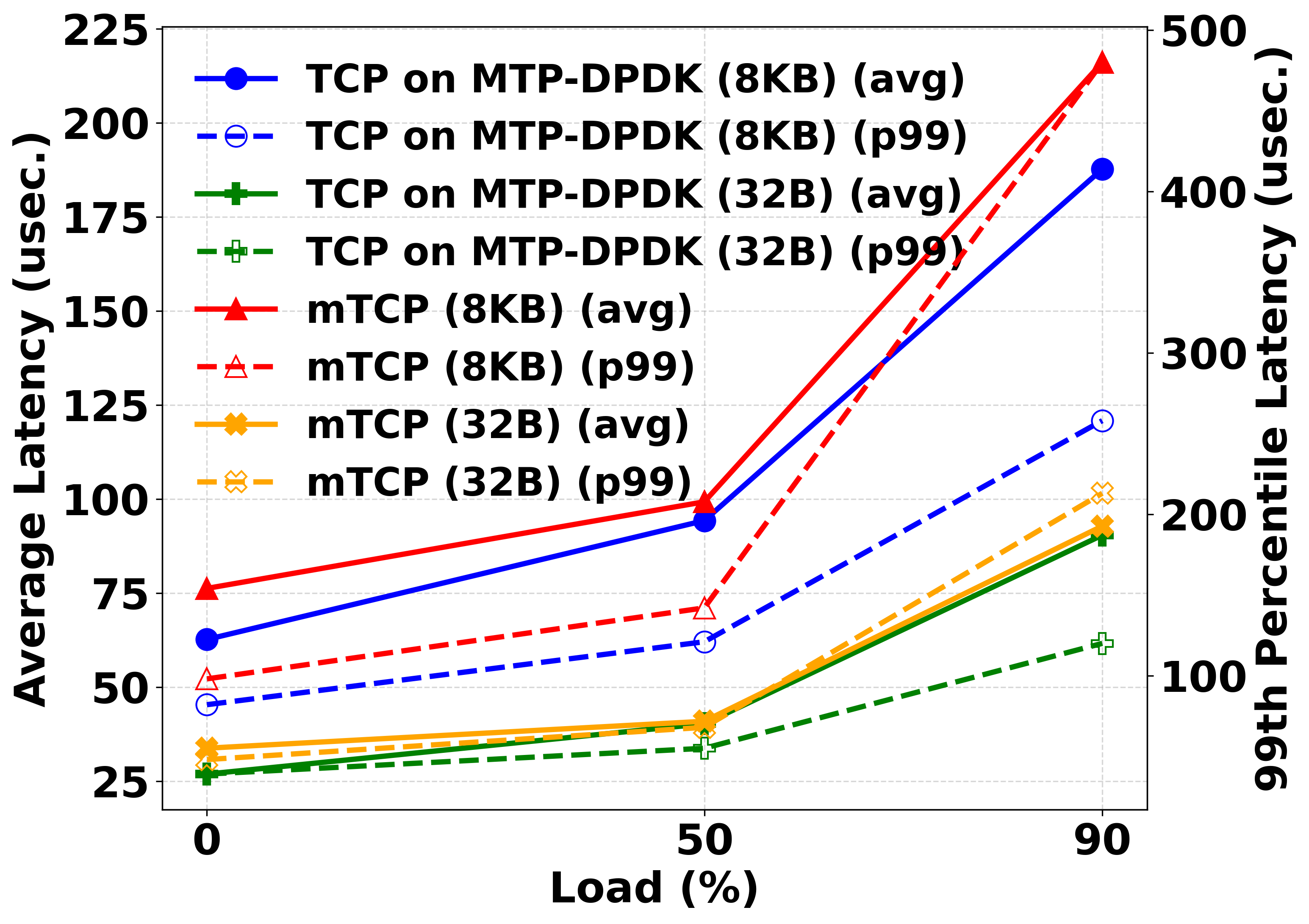}
    \caption{\small \mtpdpdk vs.\ mTCP latency}
    \label{fig:tcp_dpdk_lat}
  \end{subfigure}
\end{minipage}%
\hfill%
\begin{minipage}[b]{0.33\textwidth}
  \centering
  \begin{subfigure}[b]{\linewidth}
    \centering
    \footnotesize
    \begin{tabular}{@{} l c c @{}}
      \multicolumn{2}{r}{\textbf{Homa (\mtpxdp)}} & \textbf{Homa (eTran)} \\
      \midrule
      32B avg.\ latency ($\mu$s) & 8.45 & 8.29 \\
      1MB throughput (Gbps)      & 19.76 & 20.52 \\
      \bottomrule
    \end{tabular}

    \vspace{6pt}

    \includegraphics[width=\linewidth,height=0.0925\textheight,keepaspectratio]{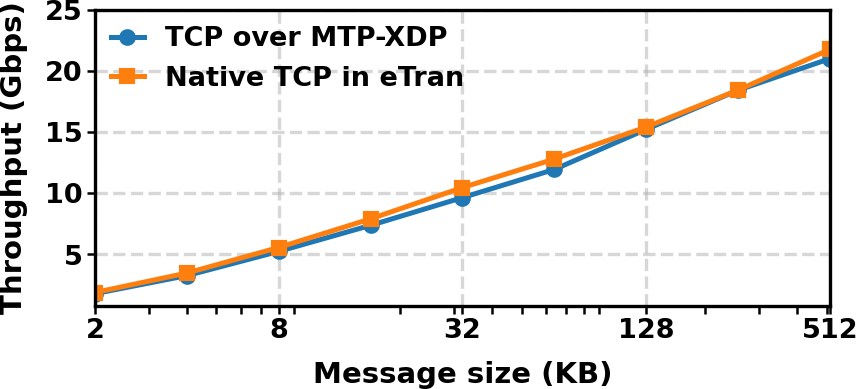}
    \caption{\small \mtpxdp vs. eTran throughput and latency}
    \label{fig:xdp-homa}
  \end{subfigure}
\end{minipage}

\par\vspace{7pt}
\caption{
\mtp-compatible stacks programmed by transport programs \textbf{achieve comparable performance to native protocol implementations} in mTCP and eTran on CloudLab.
(a,b) TCP on \mtpdpdk vs.\ mTCP.
(a) Server throughput for closed-loop 1MB resp., \# server cores increased until network becomes the bottleneck.
(b) Latency with a single server thread, 8KB and 32B resp., and increasing cross-traffic load.
(c) TCP and Homa on \mtpxdp vs.\ eTran with one server thread: TCP throughput for message sizes from 2KB–512KB and Homa latency and throughput for 32B and 1MB requests.
}

\label{fig:tcp-comp}
\end{figure*}

One of \mtp{}'s main benefits is separation of concerns and reuse (\sref{sec:motivation}). We realize this benefit by building \mtp{}-compatible stacks (\sref{sec:targets}), quantifying performance overhead against manually developed fixed protocol-stack pairings in software (\sref{sec:soft-perf}) and timing, resource overhead and microbenchmarks in our FPGA stack (\sref{sec:hw-perf}), and demonstrating separation of concern and reuse by showing that MTP-compatible stacks can run different protocols.

\subsection{Building \mtp{}-Compatible Stacks}
\label{sec:targets}

\textbf{What is an MTP-compatible stack?} An \mtp-compatible stack should provide (1) an architecture file that describes, for each of \mtp{}'s components in \sref{sec:event-processing} and \sref{sec:instructions}, i.e., events, contexts, processing logic and transport instructions, the stack's constraints and capabilities, (2) a compiler that can generate code for \mtp{} programs fitting those constraints that builds into the stack, and (3) implementations for the supported transport instructions.
In our prototype \mtp-compatible stacks described below, there is one shared compiler front-end, and each stack has its own backend compiler to generate code that it then compiles with the rest of the stack to produce the final protocol implementation in that stack.

\signpost{Software stacks: MTP-XDP and MTP-DPDK}
Our two software stacks are refactors of two existing stacks with very different execution models. \mtpxdp{} refactors eTran \cite{chen2025etran}, which implements TCP and Homa, splitting protocol execution across user space and three eBPF hooks, and \mtpdpdk{} refactors mTCP \cite{mtcp}, a TCP stack with per-core threads busy-polling the NIC through DPDK \cite{dpdk}.
While this is a substantial effort, it is one-time, concentrated in implementing the interface where many of the ``donor'' stack's core mechanisms stay as they are.
Once the stacks became \mtp-compatible, the corresponding compilers could deploy \mtp programs from different protocols on them (\sref{sec:prog}).
We provide intuition about how we performed this refactoring to build the \mtp-compatible versions and their compilers using two concrete non-trivial examples, packet generation and state consistency.

\signpost{Example: Packet generation}
Stacks specialize packet generation strategies based on various conditions, such as what triggers the packet and whether it carries data. Such conditions are either directly stated in the \mtp{} program or can be extracted by a compiler using standard static program analysis, and as such, allow the stacks to apply those strategies through specialized implementations of \mtp{}'s instruction set.

In eTran, for example, packet generation, i.e., packet construction, payload access, and transmission, is split across eBPF hooks, fast user-space data path, and a slower user-space control path. There are six different conditions with different packet generation strategies depending on which of these components the processing starts in, e.g., a duplicate ACK arriving at an eBPF hook vs.\ an upper-layer call in the data path, and whether payload must be fetched from the send buffer (details in \sref{app:stack-decisions}).
So, in building \mtpxdp, we implemented the \emph{pkt\_gen} instruction once for each of the above strategies, and our compiler extracts the conditions from the program to pick among the six and generate the right packet-generation code for each \emph{pkt\_gen} call.

mTCP, in contrast, has a single batch strategy: in each iteration of its busy loop, it generates packets for flows with pending control packets, acks, and data, in that priority order. So, in building \mtpdpdk{}, we kept this loop for the \emph{pkt\_gen} implementation: event processing marks flows that issued \emph{pkt\_gen} instructions by generic priority levels rather than specific TCP packets, and the loop constructs packets from their blueprints. This generalization also revealed an optimization opportunity discussed in \sref{sec:soft-perf}.

\signpost{Example: State consistency} Similarly, stacks specialize synchronization mechanisms for transport state based on which components share what state, whether they read or write, or whether it is global or more granular. In eTran, for example, shared state across eBPF hooks and user-space is managed by lock-free data structures or locks depending on access patterns, per-connection and per-RPC state is sharded across cores, and global state uses locks and atomics.
Our \mtpxdp compiler can extract such conditions from context declarations, key construction in parsers, and processing logic in the program  and pick the right mechanism for each context instance and variable once functionality is placed across components. More details and mTCP are discussed in \sref{app:stack-decisions}.

\signpost{Hardware stack: MTP-FPGA}
Unlike software architecture, hardware architecture, specifically what is fixed and what is programmable, is fixed at design time, and existing hardware transport stacks vary in where they draw this line.
For example, Falcon fixes the protocol logic in the data path and sends a predefined set of signals to tightly integrated on-chip cores, which can be reprogrammed for time-sensitive control decisions such as pacing rates, window sizes, and timeout values~\cite[Table 3]{falcon}. Tonic and NanoTransport make the data path programmable, but build protocol-specific assumptions about events, processing, and state access into their architectures \cite{tonic, nano}

\mtp{}'s  protocol-stack boundary not only enables explicitly stating such constraints in architecture files, it also offers new architectural design points for drawing that line.
\mtpfpga{} is a preliminary prototype of an architecture designed against the \mtp{} interface (\pcite{pita}, also Figure~\ref{fig:pita-overview} in \sref{app:fpga-ple}).
Specifically, it places functionality that is not expressed inside transport programs, i.e., event scheduling, context storage, instruction execution modules in a Verilog shell that has no protocol logic in it and is designed and optimized once.
What is stated in the program as declarations, e.g., event and context sizes, packet formats, parsing, and segmentation rules is used to reconfigure the shell through parameters.
The processing functions in the program, which do most of the protocol's online decision making, are converted to C++, pipelined with high-level synthesis (HLS), and plugged into the shell.
Notably, these functions are scoped and regular enough for HLS (\sref{sec:processing}). Full stacks have been written in HLS at high speed \cite{easynet}, but as hardware-conscious code tailored to one protocol throughout. Here, HLS only has to pipeline the program's processing functions.

\subsection{Software Performance}
\label{sec:soft-perf}

\begin{figure*}[t]
\centering
\setlength{\abovecaptionskip}{2pt}
\begin{subfigure}[b]{0.235\textwidth}
  \centering
  \includegraphics[width=\linewidth]{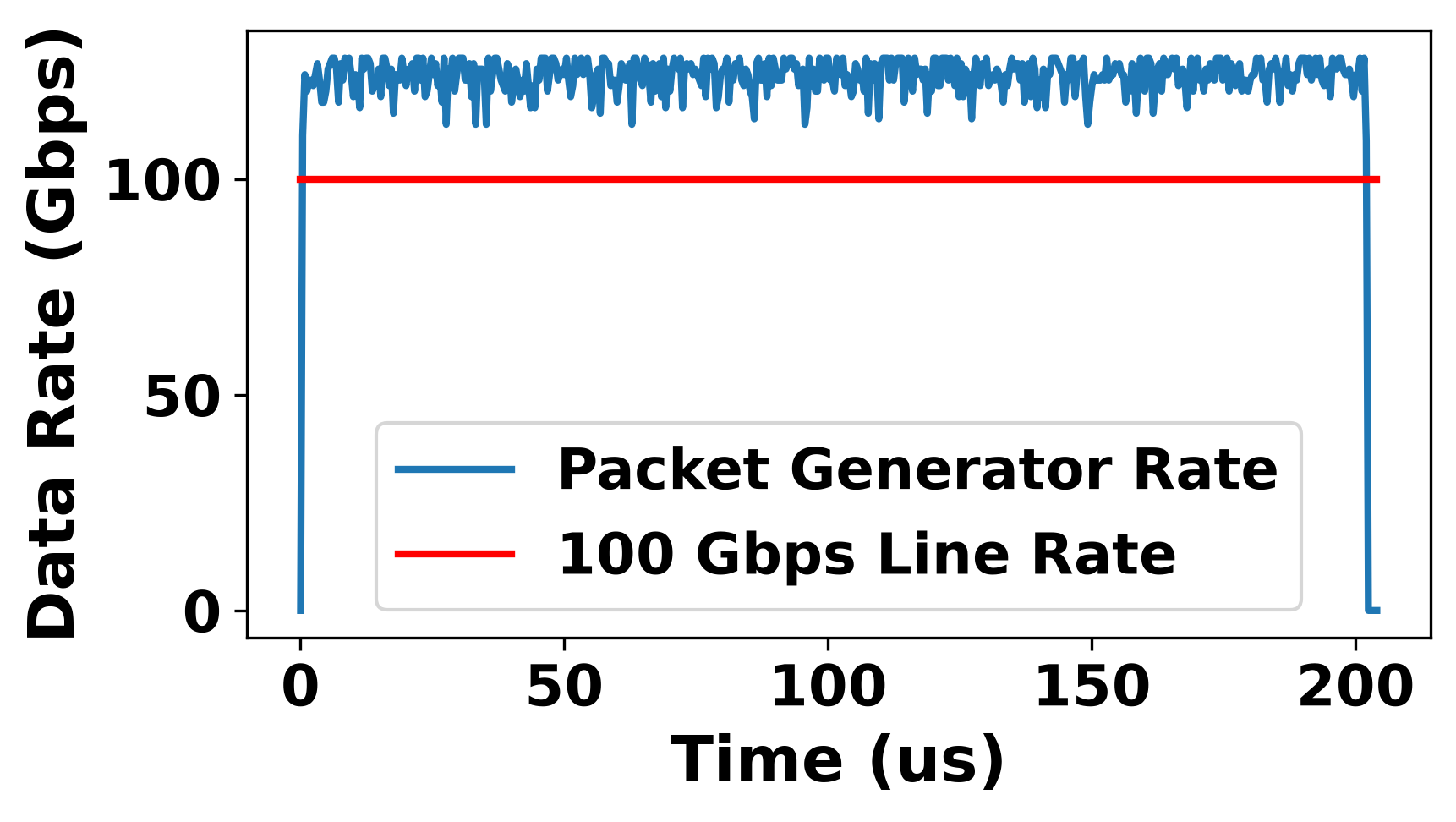}
  \caption{Packet generation, random sizes}
  \label{fig:fpga-pktgen-mixed}
\end{subfigure}\hfill
\begin{subfigure}[b]{0.235\textwidth}
  \centering
  \includegraphics[width=\linewidth]{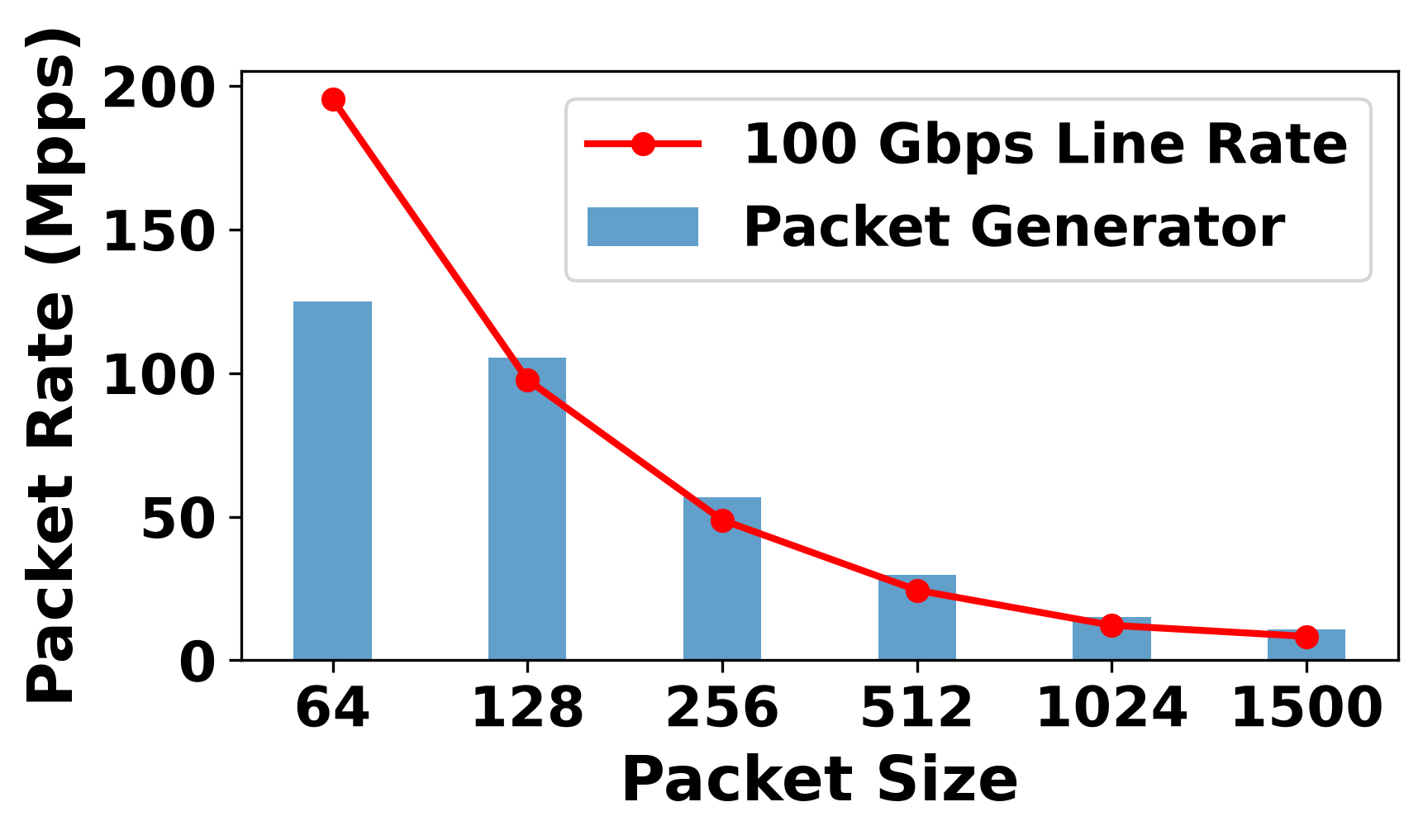}
  \caption{Packet generation, fixed sizes}
  \label{fig:fpga-pktgen-fixed}
\end{subfigure}\hfill
\begin{subfigure}[b]{0.235\textwidth}
  \centering
  \includegraphics[width=\linewidth]{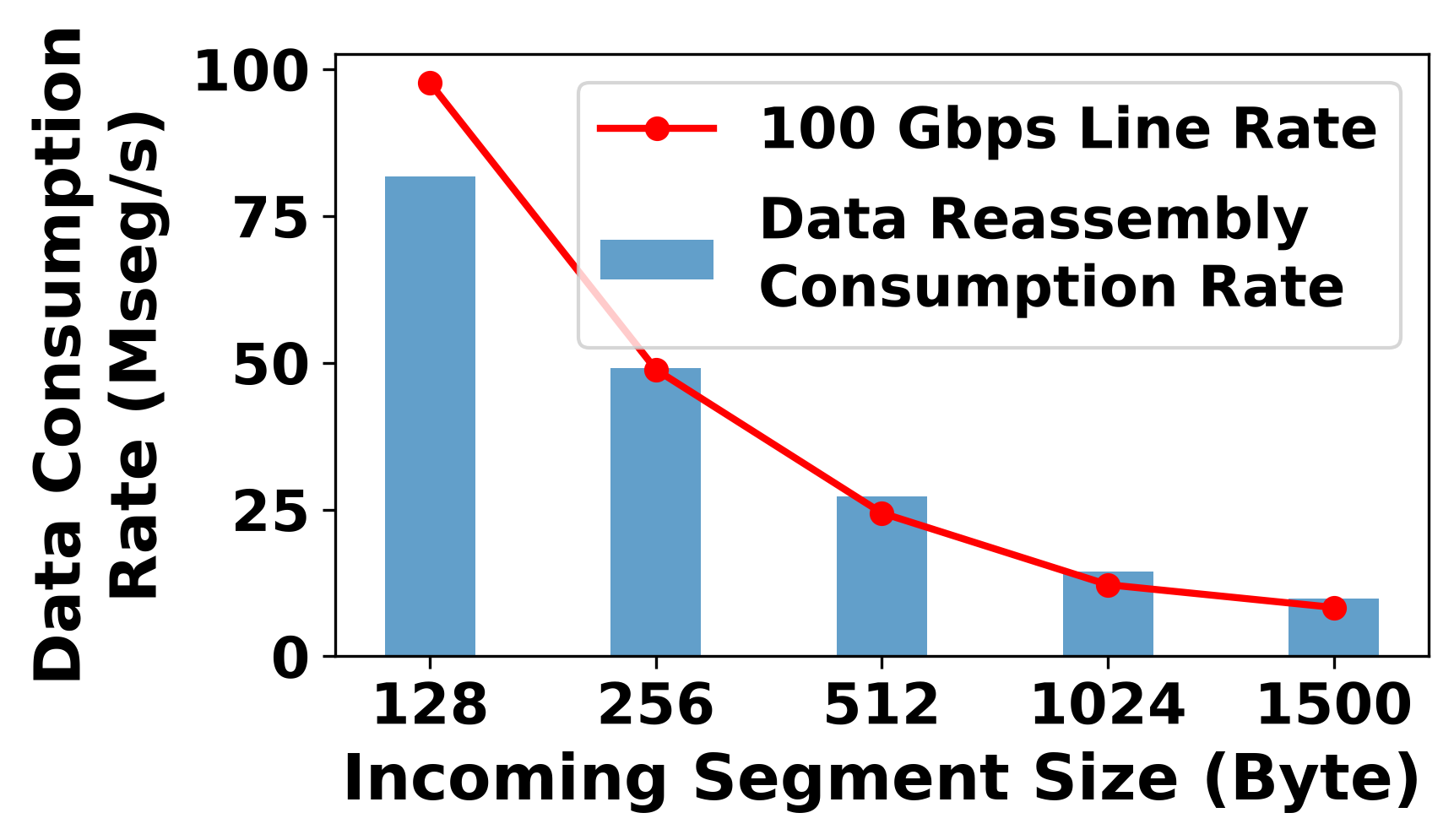}
  \caption{Data reassembly, fixed sizes}
  \label{fig:fpga-reassembly}
\end{subfigure}\hfill
\begin{subfigure}[b]{0.26\textwidth}
  \centering
  \scriptsize
  \setlength{\tabcolsep}{2pt}
  \begin{tabular}{@{}rccc@{}}
  \toprule
  \textbf{Flows} & \textbf{LUTs} & \textbf{FFs} & \textbf{BRAMs} \\
  \midrule
  128  & 18K (1.1\%) & 29K (0.8\%) & 235 (8.7\%) \\
  256  & 24K (1.4\%) & 35K (1.0\%) & 343 (12.8\%) \\
  512  & 38K (2.2\%) & 47K (1.3\%) & 527 (19.6\%) \\
  1024 & 60K (3.5\%) & 72K (2.1\%) & 898 (33.4\%) \\
  \bottomrule
  \end{tabular}
  \vspace{2pt}
  \caption{Resource utilization across flow counts}
  \label{tab:fpga-resources}
\end{subfigure}
\setlength{\belowcaptionskip}{-5pt}
\caption{\mtpfpga{} sustains line rate in its protocol-agnostic components and meets timing at 250MHz with modest resource overhead.}
\label{fig:fpga}
\vspace{-0.1in}
\end{figure*}

\signpost{Isolating the cost of \mtp{}-compatibility}
We write \mtp{} programs representing protocol decisions in mTCP's TCP and eTran's TCP and Homa, deploy them on \mtpdpdk{} and \mtpxdp{} respectively, and compare against the donor's own manual implementation. mTCP's and eTran's TCPs make different protocol decisions (\sref{sec:benefit2}), so each comparison uses the program that states its own donor's protocol decisions. This keeps the stack and protocol ``constant'' so that the performance difference in the experiments can represent the cost of making the stack \mtp{}-compatible.

\signpost{Cloudlab setup} We use three CloudLab~\cite{cloudlab} xl170 physical machines (one server, two clients), each with two 10-core Intel E5-2640v4 CPUs (2.4GHz), 64GB memory, and a ConnectX4 25 Gbps NIC, connected via Mellanox 2410 switch.

\signpost{TCP on \mtpdpdk vs mTCP \cite{mtcp}}
We use mTCP's epoll-based HTTP client and server.
To evaluate throughput, multiple client threads send back-to-back requests for a 1MB file (closed loop) to saturate the server. We increase the number of server cores until the network becomes the bottleneck and compare the server's throughput under mTCP and \mtpdpdk (Figure~\ref{fig:tcp-dpdk-throughput}).
To evaluate latency, we use one server thread, one client generating cross traffic with HTTP requests for a large 1MB file, and a second client issuing back-to-back requests for a small file in a closed loop.
We compare average and 90th-percentile request–response latency under cross-traffic load of 0\%, 50\%, and 90\% of the server thread's saturation point and file sizes of 32B and 8KB (Figure~\ref{fig:tcp_dpdk_lat}).

\mtpdpdk performs better: it reaches 21.5Gbps for two cores compared to mTCP's 18.9Gbps, and consistently provides lower latency, improving tail latency by 14\% to 46\%.
This improvement stems from a blueprint-level optimization revealed when refactoring mTCP's packet generation machinery to create \mtpdpdk{} (\sref{sec:targets}).
mTCP computes payload addresses in the generation loop that runs after event processing. For that, it takes a lock since the application thread may be modifying the send buffer concurrently, e.g., to add data in response to a send call.
In \mtpdpdk, blueprints carry everything needed to construct the packet, including payload addresses, and are built during event processing, when the event processor already holds the necessary locks.
This allows the packet generator to populate payloads without additional synchronization (\sref{app:stack-decisions}).
 While a similar optimization could in principle be engineered into mTCP, it naturally emerged from \mtp{}’s explicit packet-generation abstraction and separation of protocol semantics from execution machinery.

\signpost{TCP and Homa on \mtpxdp vs. eTran \cite{chen2025etran}}
We use eTran’s echo server setup with a single server thread and two client threads issuing back-to-back fixed-size requests, with the server replying with 100B messages.
Figure~\ref{fig:xdp-homa} shows that \mtpxdp closely tracks eTran across message sizes of 2-512KB, achieving 92\%–99\% of eTran’s throughput.
We also evaluate Homa by compiling the Homa \mtp program -- which uses the same \mtp language constructs as TCP, but with different parameters and event logic -- to the same target, \mtpxdp.
Following eTran’s evaluation, a client issues back-to-back 32B RPCs (up to two outstanding), and a single-threaded server responds with 32B replies.
The table in Figure~\ref{fig:xdp-homa} shows that \mtpxdp achieves performance comparable to eTran’s manual Homa implementation, with a 1\% increase in latency and a 3\% reduction in throughput.

The small performance gap between \mtpxdp and eTran stems primarily from an additional map lookup in certain packet generation cases (\sref{sec:targets}).
For application-triggered data transmission, events originate and data is segmented in user-space, while the event processor that determines packet-generation parameters executes in XDP-EGRESS.
\mtpxdp carries blueprint information in the first segment and stores it in an eBPF map to use for populating headers of the next segments, introducing an additional map lookup (\sref{app:stack-decisions}). More aggressive cross-boundary optimizations, e.g., partially executing application event processors in user space and safely sharing intermediate state with XDP-EGRESS, could reduce this overhead. We leave these refinements to future work.

\begin{figure*}[t]
\centering
\setlength{\abovecaptionskip}{2pt}
\begin{subfigure}[b]{0.29\textwidth}
  \centering
  \includegraphics[width=\linewidth]{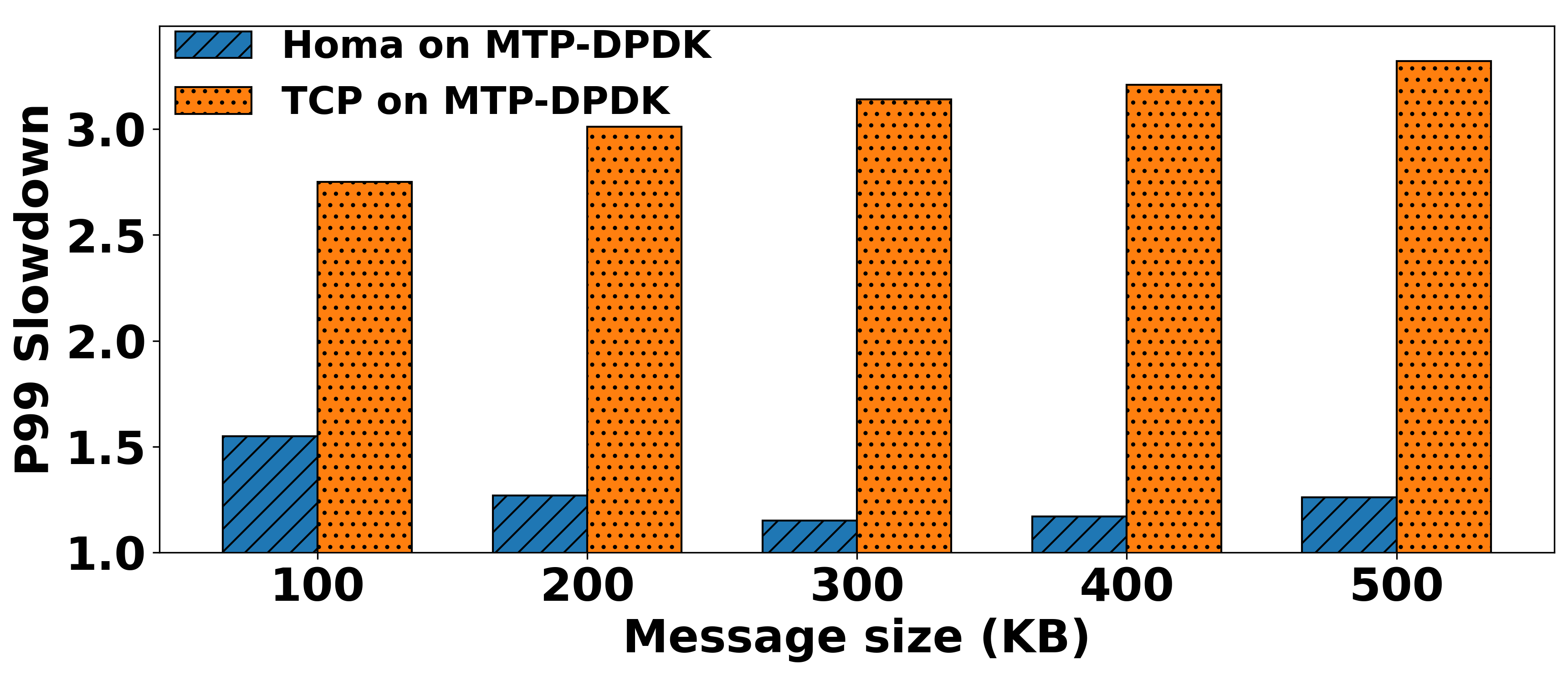}
  \caption{Homa vs.\ TCP on \mtpdpdk}
  \label{fig:dpdk-homa-tcp}
\end{subfigure}\hfill
\begin{subfigure}[b]{0.29\textwidth}
  \centering
  \includegraphics[width=\linewidth]{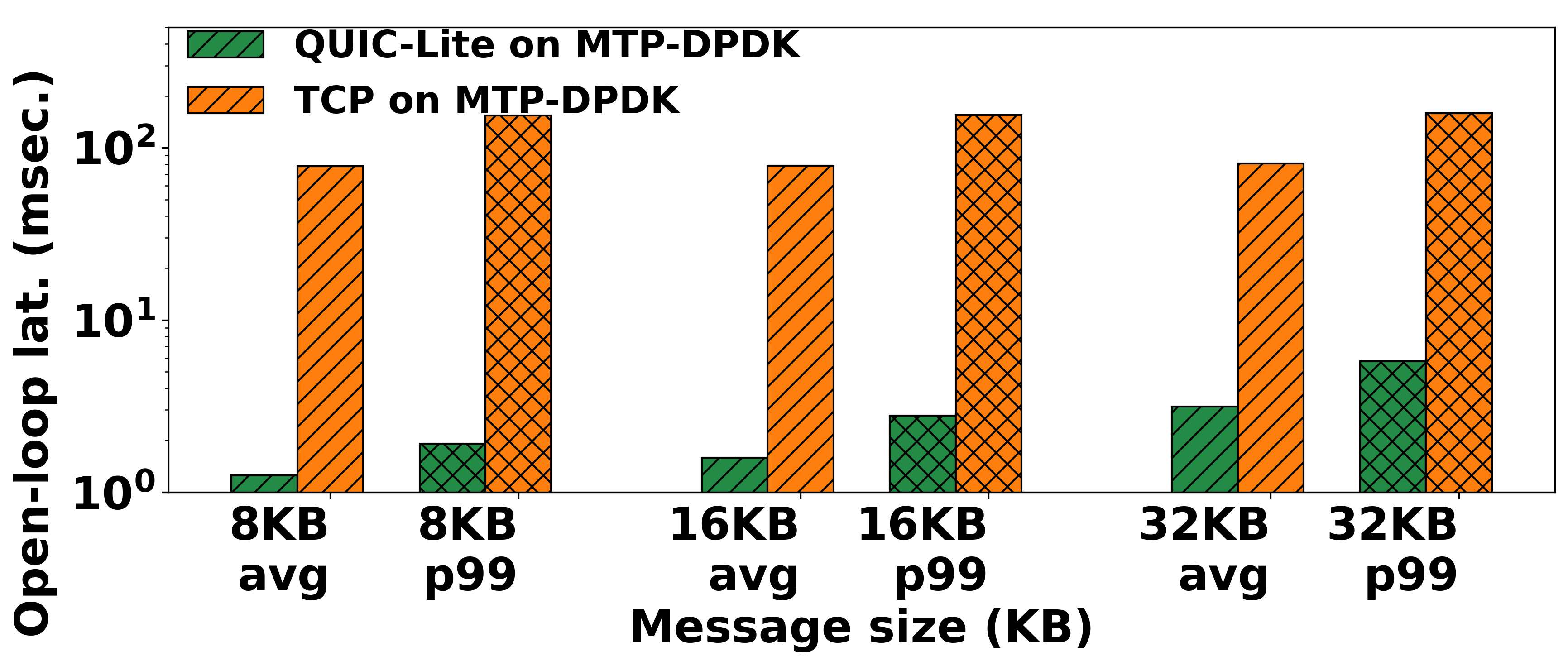}
  \caption{\quic vs.\ TCP on \mtpdpdk}
  \label{fig:dpdk-quic-tcp}
\end{subfigure}\hfill
\begin{subfigure}[b]{0.38\textwidth}
  \centering
  \scriptsize
  \setlength{\tabcolsep}{3pt}
  \begin{tabular}{@{}l cc cc@{}}
    \toprule
     & \multicolumn{2}{c}{\textbf{\mtpxdp{} (ms)}} & \multicolumn{2}{c}{\textbf{\mtpfpga{} (us)}} \\
    \cmidrule(lr){2-3} \cmidrule(l){4-5}
     & \textbf{\quic} & \textbf{TCP} & \textbf{\roce} & \textbf{TCP*} \\
    \midrule
    avg.\ latency & 3.4 & 20.1 & 20.2 & 52.4 \\
    tail latency (P90) & 5.8 & 28.7 & 56.9 & 89.4 \\
    \bottomrule
  \end{tabular}
  \vspace{2pt}
  \caption{\quic vs.\ TCP on \mtpxdp{}; TCP* vs.\ \roce{} on \mtpfpga{}}
  \label{fig:quic-xdp-table}
\end{subfigure}
\setlength{\belowcaptionskip}{-7pt}
\caption{We deploy \mtp programs for Homa and \quic{} on \mtpdpdk{} and \mtpxdp{} and TCP* and \roce{} on \mtpfpga{}, without modifying stack infrastructure. \emph{Performance is compared only against another protocol on the same stack, to validate faithful realization}; setups are described in the text. TCP* is the TCP program adjusted to \mtpfpga{}'s stated constraints (\sref{sec:prog}).}
\label{fig:prog}
\vspace{-0.1in}
\end{figure*}

\subsection{Hardware Performance}
\label{sec:hw-perf}

\signpost{Timing and resource overhead}
The dominant contributor to resource consumption in \mtpfpga{} is BRAM for per-flow event buffers, context, pending instruction buffers in instruction modules.
To evaluate hardware cost and timing characteristics, we synthesize \mtpfpga{} on an AMD Alveo U250 FPGA for 128-1024 flows, per-flow event buffers of size 32, and per-flow instruction buffer size of 8, and the TCP program described in \sref{sec:prog}.
All configurations meet timing at 250MHz, and Table~\ref{tab:fpga-resources} summarizes the resource utilization.
Despite supporting protocol-agnostic execution, \mtpfpga{} incurs modest overhead, using $\leq$3.5\% of LUTs and $\sim$2\% of FFs.
BRAM usage ranges from 9\% for 128 flows to $\sim$34\% for 1024.
The higher usage of BRAM compared to other resources reflects the stateful nature of transport processing and is also observed by prior work \cite{tonic, F4T}.

\signpost{Microbenchmarks} \mtpfpga{}'s packet generation and data reassembly modules execute the respective instructions issued from event processing pipelines. Both are protocol-agnostic and instruction driven, i.e., they make no assumptions about the protocol and rely on the input instruction to fully specify how data is segmented and placed in packets or where an incoming segment belongs in a reassembly buffer. Sustaining line rate then requires per-instruction data prefetching and interleaving instructions by their parameters (details in \sref{app:fpga-ple}).
To see if protocol-agnostic execution can maintain high-performance, we performed microbenchmarks on both (Figure~\ref{fig:fpga}). For the packet generation module, every instruction requests a single segment, which is a more demanding condition than typical workloads where one instruction often corresponds to multiple packets.
Packet generation sustains 100Gbps line rate for random packet sizes of 64B--1500B and for fixed sizes of 128B and above. Data reassembly sustains it for segment sizes of 256B and above.
We discuss in \sref{app:fpga-ple} how the gaps to line rate for very small back-to-back single packet instructions, which are rare in practice, are prototype artifacts and not fundamental limitations of the approach.

\subsection{Programmability and expressiveness}
\label{sec:prog}

We deploy \mtp{} programs for TCP, Homa, and \quic{} (a lightweight adaptation of QUIC) on our software stacks, \mtpdpdk{} and \mtpxdp{}, including protocols their donors, mTCP and eTran, never implemented. We also deploy TCP and \roce{} programs on \mtpfpga{}, adjusting the TCP program to the constraints stated in \mtpfpga{}'s architecture file. We validate every deployment against a TCP program on the same stack, in scenarios chosen to show each protocol's distinct behavior (Figure~\ref{fig:prog}). We show (1) separation of concerns and reuse: new protocols run on an \mtp{}-compatible stack without re-engineering its infrastructure, and (2) expressiveness by covering protocols of different kinds and stacks in software and hardware with their own constraints.

\signpost{The programs}
The TCP, Homa, and \quic{} programs are 753, 1205, and 920 lines of \mtp{}, precisely capturing protocol decisions (R1)  while staying an order of magnitude below the $\sim$15K lines of stack code.
The TCP and Homa programs reflect the protocol decisions of the donor stacks' implementations, not a simplified subset (For an RFC-complete TCP program, see \sref{sec:benefit3}). \quic{} connections have multiple streams and its packets can carry frames from different streams. The \mtpfpga{} TCP and RoCEv2 programs additionally adjust some decisions to the constraints in its architecture file, as discussed later in the section.

\signpost{Reprogramming \mtpdpdk}
Homa and \quic{} are fundamentally different protocols from the TCP implemented in mTCP, and both run on \mtpdpdk{}'s generalized machinery (\sref{sec:targets}).
The illustrative scenario demonstrating Homa's deployment on \mtpdpdk{} is shown in Figure~\ref{fig:dpdk-homa-tcp}.
One client issues back-to-back RPC requests of 100–500KB, while a second client sends 1MB requests to generate 50\% load on a single-threaded server; the server responds with 170B replies.
Homa’s unscheduled bytes are set to 60KB, ensuring that its receiver-driven grant mechanism is exercised and that short RPCs are prioritized over long ones.
Homa on \mtpdpdk demonstrates expected behavior of maintains slowdown close to one against TCP's slowdown of  $>3$.

The illustrative scenario demonstrating \quic{}'s deployment on \mtpdpdk is shown in Figure~\ref{fig:dpdk-quic-tcp}.
A client issues alternating requests for small (8–32KB) and large (1MB) objects over a single connection to a single-threaded server.
In \quic, short and long requests are placed on separate streams switching between them every 100KB for packet generation, whereas TCP multiplexes them over a single byte stream.
Accordingly, \quic on \mtpdpdk demonstrates expected behavior of much lower latency for short messages than TCP.

\signpost{Reprogramming \mtpxdp and portability}
We deploy the \quic \mtp program on \mtpxdp to demonstrate both programmability -- supporting a protocol not previously implemented in eTran -- and portability, by reusing the same \mtp program on a substantially different target.
We use a similar scenario as above with eTran’s echo server application: a client sends 32KB and 1MB messages in an open-loop, round-robin fashion over a single connection.
Figure~\ref{fig:quic-xdp-table} shows that, as expected, \quic on \mtpxdp has much lower latency for short messages than TCP.

\signpost{Reprogramming \mtpfpga{} under stated constraints}
\mtpfpga{} is a more constrained stack than others. For example, its reassembly module only handles a bounded number of gaps in the reassembly buffer and its parser and packet generator only handle fixed header layouts.
These constraints are stated against the \mtp{} abstraction in its architecture file.
Given these explicit constraints, we adjusted the TCP program to drop segments that create more gaps than the reassembly module can handle, and removed options from its blueprint, so it computes RTTs from sender-side measurements.

The modified TCP (TCP*) as well as our \roce{} program, can be deployed on the same hardware architecture, just by reconfiguring the Verilog shell and plugging in the HLS output for the event processing functions (\sref{sec:targets}, see \sref{app:fpga-ple} for HLS output details).
Figure~\ref{fig:prog}(c) shows the two protocols' distinct behavior on the same datapath using a simple key-value client-server application in cycle-accurate hardware simulation.
The client issues back-to-back requests with 4B keys at 8M Req/s, and the server responds with 64B values.
The round-trip time is set to 6.4$\mu$s, and the queue size is 24KB (400 minimum-sized packets). For \roce, the ECN marking threshold for DCQCN is set to 3KB ($\sim$ 50 minimum-sized packets).
We induce congestion by temporarily reducing the queue drain rate to half the request generation rate, causing queue buildup before restoring the original rate.
As expected, \roce{} with DCQCN keeps average and tail latency well below TCP's.

\section{Benefit 2: Communicating intended behavior}
\label{sec:benefit2}
\begin{figure*}[t]
\centering
\setlength{\abovecaptionskip}{1pt}
\setlength{\belowcaptionskip}{-15pt}
\begin{minipage}[c]{0.43\textwidth}
  \centering
  \scriptsize
  \setlength{\tabcolsep}{3pt}
  \renewcommand{\arraystretch}{1.25}
  \begin{tabular}{@{}ll@{}}
  \toprule
  \textbf{Interop problem} & \textbf{Consequence} \\
  \midrule
  mTCP window scale 7, eTran silently 3 & 18$\times$ lower throughput \\
  eTran ignores data-carrying dup ACKs & no fast retransmit \\
  eTran never re-ACKs data mTCP retransmits & connection stalls on ACK loss \\
  \bottomrule
  \end{tabular}
\end{minipage}\hfill
\begin{minipage}[c]{0.56\textwidth}
  \centering
  \begin{tikzpicture}[
    font=\scriptsize,
    box/.style={draw, align=center, inner sep=2.5pt, fill=white},
    arr/.style={->, thick}
  ]
  \node[box] (ev) {event\\struct};
  \node[box, right=2.5mm of ev] (ctx) {context};
  \node[box, right=2.5mm of ctx] (proc) {event\\processors};
  \draw[arr] (ev) -- (ctx);
  \draw[arr] (ctx) -- (proc);
  \node[draw, dashed, inner sep=2pt, fit=(ev)(ctx)(proc)] (group) {};
  \node[box, above=4.5mm of ev.north west, anchor=south west] (evlist) {event\\list};
  \node[box, right=3mm of evlist, text width=100pt, align=center] (contracts) {cross-event contracts: names, parameters, dependency map};
  \draw[arr] (evlist) -- (contracts);
  \draw[arr] (evlist.south) -- (evlist.south|-group.north) node[midway, right, font=\tiny\bfseries] {$\times$12};
  \node[box, right=4mm of group] (asm) {assemble\\$+$ compile};
  \draw[arr] (group) -- (asm);
  \node[draw, thick, inner sep=3.5pt, fit=(evlist)(contracts)(group)(asm),
        label={[font=\scriptsize\bfseries, yshift=-1mm]above:AI agent pipeline}] (agent) {};
  \coordinate (rfcNE) at ($(agent.north west)+(-4mm,0)$);
  \coordinate (rfcSW) at ($(agent.south west)+(-12mm,0)$);
  \node[box, fill=blue!10, inner sep=0pt, fit=(rfcNE)(rfcSW)] (rfc) {};
  \node[rotate=90] at (rfc.center) {\shortstack{\textbf{RFC 9293} $+$\\referenced RFCs}};
  \draw[arr] (rfc.east|-evlist.west) -- (evlist.west);
  \node[box, anchor=north] (gen) at ($(agent.north east)+(14mm,-3.5pt)$) {generated\\TCP};
  \node[box, anchor=south] (nat) at ($(agent.south east)+(14mm,3.5pt)$) {native\\TCP};
  \draw[<->, thick] (gen) -- (nat) node[midway, right=1pt, font=\scriptsize\bfseries] (ilbl) {interop};
  \node[draw, thick, inner sep=3pt, fit=(gen)(nat)(ilbl),
        label={[font=\scriptsize\bfseries, yshift=-1mm]above:NS3}] (ns3) {};
  \draw[arr] (asm.east) -- ([xshift=3.5mm]asm.east) |- (gen.west);
  \begin{scope}[on background layer]
    \node[draw=none, fill=orange!10, inner sep=3.5pt, fit=(evlist)(contracts)(group)(asm)] {};
    \node[draw=none, fill=green!10, inner sep=3pt, fit=(gen)(nat)(ilbl)] {};
  \end{scope}
  \end{tikzpicture}
\end{minipage}
\setlength{\belowcaptionskip}{-5pt}
\caption{(Left) interoperability issues found from mTCP's and eTran's \mtp{} programs (\sref{sec:benefit2}); (right) agentic TCP generation from RFCs (\sref{sec:benefit3}).}
\label{fig:interop-agentic}
\vspace{-0.1in}
\end{figure*}

A transport program describes, precisely, the protocol behavior a stack implements, whether that stack is \mtp{}-compatible or a fixed protocol-stack pairing. This supports testing, verification, and comparing two stacks' protocol behavior for divergence (\sref{sec:motivation}). We demonstrate the last one: we found and fixed interoperability issues between mTCP and eTran through their \mtp{} programs alone, which were used in \sref{sec:soft-perf} and we refer to as \code{tcp-mtcp.mtp} and \code{tcp-etran.mtp} in this section.

We analyzed the two programs with an AI agent (Opus 5). Figure~\ref{fig:interop-agentic} (left) highlights three of the divergences we found. First, window scaling mismatch: mTCP negotiates a factor of 7 in the handshake, eTran sends no option and silently scales by 3, so the two sides misread each other's receive window, which results in significant low performance (180Mbps vs. 3.34Gbps after the fix). Second, missed duplicate acks: eTran counts a duplicate ACK only when the segment has no payload, and mTCP piggybacks data on duplicate ACKs when both sides send, so fast retransmit never triggers and recovery falls to timeouts. Third, a stall: eTran does not acknowledge duplicate data, so when mTCP's both windows are full and eTran's first ACKs are lost, mTCP retransmits data eTran already has and the connection stalls.

We acknowledge that mTCP and eTran are prototypes, built for particular environments and not with interoperability as a first goal. What we want to highlight is that the divergences were found by analyzing two \mtp{} programs -- not by running the stacks against each other and debugging from packet traces -- and, hence, each was easy to localize and fix. None of this requires the stacks to be \mtp{}-compatible. \mtp{} programs written to capture the protocol behavior of fixed stacks are enough. Automating this comparison in a sound and efficient way, with AI agents or formal approaches, is far more feasible over two transport programs than over two codebases, and along with automatic derivation of \mtp{} programs from fixed stacks, are promising research directions.

\section{Benefit 3: Structure for agentic development}
\label{sec:benefit3}

AI agents work best on bounded tasks with checkable outputs, and \mtp{} gives transport development that structure. We demonstrate it with an agentic pipeline that generates a TCP implementation from RFC 9293 and the relevant RFCs it points to, i.e., a corpus of about a hundred pages. An AI model cannot implement that in one shot.
\mtp{} provides the structure to break up the task into an agentic flow. The pipeline (Figure~\ref{fig:interop-agentic}, right) first extracts the protocol's event list from the RFCs and generates a shared cross-event contract containing a registry of variable and parameter names and a map of the shared context fields and timers. It then fans out per event, twelve for TCP, generating each event's struct, context fields, and event processing functions from only that event's RFC excerpt and the contracts. Because \mtp{} is C-like, the pipeline emits constrained C in \mtp{}'s shape, and a final stage assembles the pieces and compiles them as a check.

We port the generated TCP into our MTP-compatible NS3 module and run it against NS3's native TCP over a simulated link, across twenty interoperability cases covering connection management, data transfer, and loss recovery, including half-close, out-of-order arrival, and data-carrying ACKs.
The interoperability suite is, of course, not a proof of correctness, and we checked the generated MTP program manually to ensure both RFC and MTP compliance. Our goal here is to show the quality of what \mtp{}'s structure lets an agent produce as well as the expressiveness of the fixed core instruction set of \mtp{}.
Finally, this is yet another demonstration of \mtp{}'s expressiveness (\sref{sec:prog}): a program that captures a protocol at the RFC level.

\section{Related Work}

We have discussed the work closest to \mtp{} in \sref{sec:statusquo}, and programmable hardware transports, Tonic, NanoTransport, and Falcon, in \sref{sec:targets}.

Network programming has been an active research area over the past 15 years, with extensive work on abstractions for stateless and stateful packet processing, most notably P4 and its ecosystem \cite{foster2011frenetic, monsanto2012compiler, soule2014merlin, arashloo2016snap, bosshart2014p4, mcclurg2016event, pontarelli2019flowblaze, hogan2022modular, doenges2021petr4, fattaholmanan2021p4, baldi2021incremental, baldi2019dapipe, yu2020mantis}. Ibanez et al.~\cite{ibanez2019event} extend P4 to support data-plane events beyond packet arrival, e.g., timers and recirculation.
These abstractions primarily target L2/L3 packet processing, whereas transport programs specify transport-layer protocols, which have fundamentally different semantics and requirements, including operating on application and payload data (\sref{sec:motivation}, \sref{sec:mtp}).

Prior work tests protocol behavior from blackbox implementation: Prognosis learns behavioral models of closed-box QUIC implementations to find spec violations and interoperability bugs \cite{prognosis}, and \cite{mcmillan-quic} build a formal, testable QUIC specification to find implementation errors and ambiguities in the then-draft standard. Eywa uses LLMs to build models of intended protocol behavior from RFCs and other natural-language sources and tests DNS, BGP, and SMTP implementations against them and model TCP handshake \cite{eywa}. Transport programs complement these approaches. They allow a stack's protocol decisions to be stated explicitly in a domain-specific language, which can be used for model learning from blackbox implementations, formal analysis, and as a precise target for models extracted from standards documents (\sref{sec:benefit3}).

\section*{Conclusion}
We proposed transport programs, written in a language we call \mtp{}, that state a protocol's decisions apart from the stack's execution machinery. They can run on \mtp{}-compatible stacks, describe protocol behavior of fixed stacks for testing, verification, and catching interoperability issues, and structure agentic transport development. We believe these are structures the community can build on and refine, opening research directions across transport stacks and protocols.

\bibliographystyle{unsrt}
\bibliography{reference}

@misc{tcp-rfc,
  title = {TCP RFC},
  howpublished = {https://datatracker.ietf.org/doc/html/rfc9293},
  note = {Accessed: January 2026} 
}

@inproceedings{quic-paper,
  title={The quic transport protocol: Design and internet-scale deployment},
  author={Langley, Adam and Riddoch, Alistair and Wilk, Alyssa and Vicente, Antonio and Krasic, Charles and Zhang, Dan and Yang, Fan and Kouranov, Fedor and Swett, Ian and Iyengar, Janardhan and others},
  booktitle={Proceedings of the conference of the ACM special interest group on data communication},
  pages={183--196},
  year={2017}
}

@inproceedings{homa,
  title={Homa: A receiver-driven low-latency transport protocol using network priorities},
  author={Montazeri, Behnam and Li, Yilong and Alizadeh, Mohammad and Ousterhout, John},
  booktitle={Proceedings of the 2018 Conference of the ACM Special Interest Group on Data Communication},
  pages={221--235},
  year={2018}
}

@inproceedings{roce,
  title={RDMA over commodity ethernet at scale},
  author={Guo, Chuanxiong and Wu, Haitao and Deng, Zhong and Soni, Gaurav and Ye, Jianxi and Padhye, Jitu and Lipshteyn, Marina},
  booktitle={Proceedings of the 2016 ACM SIGCOMM Conference},
  pages={202--215},
  year={2016}
}

@inproceedings{xdp-paper,
  title={The express data path: Fast programmable packet processing in the operating system kernel},
  author={H{\o}iland-J{\o}rgensen, Toke and Brouer, Jesper Dangaard and Borkmann, Daniel and Fastabend, John and Herbert, Tom and Ahern, David and Miller, David},
  booktitle={Proceedings of the 14th international conference on emerging networking experiments and technologies},
  pages={54--66},
  year={2018}
}

@misc{p4,
    title = {{P4 Open Source Programming Language}},
    howpublished = {https://p4.org/},
    note = {Accessed: January 2026}
}

@inproceedings{tonic,
  title={Enabling Programmable Transport Protocols in $\{$High-Speed$\}$$\{$NICs$\}$},
  author={Arashloo, Mina Tahmasbi and Lavrov, Alexey and Ghobadi, Manya and Rexford, Jennifer and Walker, David and Wentzlaff, David},
  booktitle={17th USENIX Symposium on Networked Systems Design and Implementation (NSDI 20)},
  pages={93--109},
  year={2020}
}

@inproceedings{sivaraman2016packet,
  title={Packet transactions: High-level programming for line-rate switches},
  author={Sivaraman, Anirudh and Cheung, Alvin and Budiu, Mihai and Kim, Changhoon and Alizadeh, Mohammad and Balakrishnan, Hari and Varghese, George and McKeown, Nick and Licking, Steve},
  booktitle={Proceedings of the 2016 ACM SIGCOMM Conference},
  pages={15--28},
  year={2016}
}

@inproceedings{hogan2022modular,
  title={Modular switch programming under resource constraints},
  author={Hogan, Mary and Landau-Feibish, Shir and Arashloo, Mina Tahmasbi and Rexford, Jennifer and Walker, David},
  booktitle={19th USENIX Symposium on Networked Systems Design and Implementation (NSDI 22)},
  pages={193--207},
  year={2022}
}

@article{doenges2021petr4,
  title={Petr4: formal foundations for p4 data planes},
  author={Doenges, Ryan and Arashloo, Mina Tahmasbi and Bautista, Santiago and Chang, Alexander and Ni, Newton and Parkinson, Samwise and Peterson, Rudy and Solko-Breslin, Alaia and Xu, Amanda and Foster, Nate},
  journal={Proceedings of the ACM on Programming Languages},
  volume={5},
  number={POPL},
  pages={1--32},
  year={2021},
  publisher={ACM New York, NY, USA}
}

@inproceedings{arashloo2016snap,
  title={SNAP: Stateful network-wide abstractions for packet processing},
  author={Arashloo, Mina Tahmasbi and Koral, Yaron and Greenberg, Michael and Rexford, Jennifer and Walker, David},
  booktitle={Proceedings of the 2016 ACM SIGCOMM Conference},
  pages={29--43},
  year={2016}
}

@inproceedings{fattaholmanan2021p4,
  title={P4 weaver: Supporting modular and incremental programming in P4},
  author={Fattaholmanan, Ali and Baldi, Mario and Carzaniga, Antonio and Soul{\'e}, Robert},
  booktitle={Proceedings of the ACM SIGCOMM Symposium on SDN Research (SOSR)},
  pages={54--65},
  year={2021}
}

@misc{baldi2021incremental,
  title={Incremental development of a data plane program},
  author={Baldi, Mario},
  year={2021},
  month=may # "~11",
  publisher={Google Patents},
  note={US Patent 11,005,754}
}

@inproceedings{baldi2019dapipe,
  title={DaPIPE a data plane incremental programming environment},
  author={Baldi, Mario},
  booktitle={2019 ACM/IEEE Symposium on Architectures for Networking and Communications Systems (ANCS)},
  pages={1--6},
  year={2019},
  organization={IEEE}
}

@article{foster2011frenetic,
  title={Frenetic: A network programming language},
  author={Foster, Nate and Harrison, Rob and Freedman, Michael J and Monsanto, Christopher and Rexford, Jennifer and Story, Alec and Walker, David},
  journal={ACM Sigplan Notices},
  volume={46},
  number={9},
  pages={279--291},
  year={2011},
  publisher={ACM New York, NY, USA}
}

@article{monsanto2012compiler,
  title={A compiler and run-time system for network programming languages},
  author={Monsanto, Christopher and Foster, Nate and Harrison, Rob and Walker, David},
  journal={Acm sigplan notices},
  volume={47},
  number={1},
  pages={217--230},
  year={2012},
  publisher={ACM New York, NY, USA}
}

@article{bosshart2014p4,
  title={P4: Programming protocol-independent packet processors},
  author={Bosshart, Pat and Daly, Dan and Gibb, Glen and Izzard, Martin and McKeown, Nick and Rexford, Jennifer and Schlesinger, Cole and Talayco, Dan and Vahdat, Amin and Varghese, George and others},
  journal={ACM SIGCOMM Computer Communication Review},
  volume={44},
  number={3},
  pages={87--95},
  year={2014},
  publisher={ACM New York, NY, USA}
}

@article{mcclurg2016event,
  title={Event-driven network programming},
  author={McClurg, Jedidiah and Hojjat, Hossein and Foster, Nate and {\v{C}}ern{\`y}, Pavol},
  journal={ACM SIGPLAN Notices},
  volume={51},
  number={6},
  pages={369--385},
  year={2016},
  publisher={ACM New York, NY, USA}
}

@inproceedings{yu2020mantis,
  title={Mantis: Reactive programmable switches},
  author={Yu, Liangcheng and Sonchack, John and Liu, Vincent},
  booktitle={Proceedings of the Annual conference of the ACM Special Interest Group on Data Communication on the applications, technologies, architectures, and protocols for computer communication},
  pages={296--309},
  year={2020}
}

@inproceedings{soule2014merlin,
  title={Merlin: A language for provisioning network resources},
  author={Soul{\'e}, Robert and Basu, Shrutarshi and Marandi, Parisa Jalili and Pedone, Fernando and Kleinberg, Robert and Sirer, Emin Gun and Foster, Nate},
  booktitle={Proceedings of the 10th ACM International on Conference on emerging Networking Experiments and Technologies},
  pages={213--226},
  year={2014}
}

@inproceedings{pontarelli2019flowblaze,
  title={$\{$FlowBlaze$\}$: Stateful Packet Processing in Hardware},
  author={Pontarelli, Salvatore and Bifulco, Roberto and Bonola, Marco and Cascone, Carmelo and Spaziani, Marco and Bruschi, Valerio and Sanvito, Davide and Siracusano, Giuseppe and Capone, Antonio and Honda, Michio and others},
  booktitle={16th USENIX Symposium on Networked Systems Design and Implementation (NSDI 19)},
  pages={531--548},
  year={2019}
}

@inproceedings{narayan2018restructuring,
  title={Restructuring endpoint congestion control},
  author={Narayan, Akshay and Cangialosi, Frank and Raghavan, Deepti and Goyal, Prateesh and Narayana, Srinivas and Mittal, Radhika and Alizadeh, Mohammad and Balakrishnan, Hari},
  booktitle={Proceedings of the 2018 Conference of the ACM Special Interest Group on Data Communication},
  pages={30--43},
  year={2018}
}

@inproceedings{mtcp,
  title={$\{$mTCP$\}$: a highly scalable user-level $\{$TCP$\}$ stack for multicore systems},
  author={Jeong, EunYoung and Wood, Shinae and Jamshed, Muhammad and Jeong, Haewon and Ihm, Sunghwan and Han, Dongsu and Park, KyoungSoo},
  booktitle={11th USENIX Symposium on Networked Systems Design and Implementation (NSDI 14)},
  pages={489--502},
  year={2014}
}

@misc{dpdk,
    title = {{DPDK}},
    howpublished = {https://www.dpdk.org/},
    note = {Accessed: January 2026}
}

@misc{tldk,
    title = {{Transport Layer Development Kit (TLDK)}},
    howpublished = {https://github.com/FDio/tldk},
    note = {Accessed: January 2026}
}

@inproceedings{nano,
author = {Arslan, Serhat and Ibanez, Stephen and Mallery, Alex and Kim, Changhoon and McKeown, Nick},
title = {NanoTransport: A Low-Latency, Programmable Transport Layer for NICs},
year = {2021},
isbn = {9781450390842},
publisher = {Association for Computing Machinery},
address = {New York, NY, USA},
url = {https://doi.org/10.1145/3482898.3483365},
doi = {10.1145/3482898.3483365},
booktitle = {Proceedings of the ACM SIGCOMM Symposium on SDN Research (SOSR)},
pages = {13–26},
numpages = {14},
location = {Virtual Event, USA},
series = {SOSR '21}
}

@inproceedings{chen2025etran,
  title={$\{$eTran$\}$: Extensible Kernel Transport with $\{$eBPF$\}$},
  author={Chen, Zhongjie and Meng, Qingkai and Lao, ChonLam and Liu, Yifan and Ren, Fengyuan and Yu, Minlan and Zhou, Yang},
  booktitle={22nd USENIX Symposium on Networked Systems Design and Implementation (NSDI 25)},
  pages={407--425},
  year={2025}
}

@inproceedings{prognosis,
  title={Prognosis: closed-box analysis of network protocol implementations},
  author={Ferreira, Tiago and Brewton, Harrison and D'Antoni, Loris and Silva, Alexandra},
  booktitle={ACM SIGCOMM},
  year={2021}
}

@inproceedings{prolac,
  title={A readable TCP in the Prolac protocol language},
  author={Kohler, Eddie and Kaashoek, M Frans and Montgomery, David R},
  booktitle={SIGCOMM},
  year={1999}
}

@inproceedings{ibanez2019event,
  title={Event-driven packet processing},
  author={Ibanez, Stephen and Antichi, Gianni and Brebner, Gordon and McKeown, Nick},
  booktitle={Proceedings of the 18th ACM Workshop on Hot Topics in Networks},
  pages={133--140},
  year={2019}
}

@inproceedings{cloudlab,
  title={The design and operation of $\{$CloudLab$\}$},
  author={Duplyakin, Dmitry and Ricci, Robert and Maricq, Aleksander and Wong, Gary and Duerig, Jonathon and Eide, Eric and Stoller, Leigh and Hibler, Mike and Johnson, David and Webb, Kirk and others},
  booktitle={USENIX ATC},
  pages={1--14},
  year={2019}
}

@article{srd,
  title={A cloud-optimized transport protocol for elastic and scalable hpc},
  author={Shalev, Leah and Ayoub, Hani and Bshara, Nafea and Sabbag, Erez},
  journal={IEEE micro},
  volume={40},
  number={6},
  pages={67--73},
  year={2020},
  publisher={IEEE}
}

@inproceedings{falcon,
  title={Falcon: A reliable, low latency hardware transport},
  author={Singhvi, Arjun and Dukkipati, Nandita and Chandra, Prashant and Wassel, Hassan MG and Sharma, Naveen Kr and Rebello, Anthony and Schuh, Henry and Kumar, Praveen and Montazeri, Behnam and Bansod, Neelesh and others},
  booktitle={Proceedings of the ACM SIGCOMM 2025 Conference},
  pages={248--263},
  year={2025}
}

@techreport{UET,
  title={UEC 1.0: New High-Performance Standard for Scaling HPC-AI},
  author={{Intersect360 Research}},
  institution= {Ultra Ethernet Consortium},
  year={2025},
  month=jun,
  type= {White Paper},
  url= {https://ultraethernet.org/wp-content/uploads/sites/20/2025/06/UEC1.0Whitepaper.pdf},
}

@misc{linux_kernel_devmem,
  author       = {{Linux Kernel Development Community}},
  title        = {Device Memory TCP},
  howpublished = {\url{https://docs.kernel.org/networking/devmem.html}},
  note         = {The Linux Kernel Documentation},
  note = {Accessed: January 2026}
}

@inproceedings{homa-linux,
  author = {Ousterhout, John},
  title = {A {Linux} Kernel Implementation of the {Homa} Transport Protocol},
  booktitle = {2021 USENIX Annual Technical Conference (USENIX ATC 21)},
  pages = {99--115},
  year = {2021}
}

@inproceedings{mcmillan-quic,
  author = {McMillan, Kenneth L. and Zuck, Lenore D.},
  title = {Formal Specification and Testing of {QUIC}},
  booktitle = {Proceedings of the ACM Special Interest Group on Data Communication (SIGCOMM '19)},
  year = {2019}
}

@misc{huitema-timeouts,
  author = {Huitema, Christian},
  title = {{QUIC} Timeouts and Handshake Interop},
  howpublished = {\url{https://www.privateoctopus.com/2024/01/19/quic-timeouts-and-handshake-interop.html}},
  year = {2024}
}

@techreport{rfc5040,
  title={A Remote Direct Memory Access Protocol Specification},
  author={Recio, Renato and Metzler, Bernard and Culley, Paul and Hilland, Jeff and Garcia, Dave},
  institution={RFC 5040},
  year={2007}
}

@misc{pita,
  title={A Protocol-Independent Transport Architecture},
  author={Kimiya Mohammadtaheri and David Gao and Samuel Zhang and Matthew Chen and Eric Su and Pengyu Ji and Saad Syed and Chris Neely and Mario Baldi and Nachiket Kapre and Mina Tahmasbi Arashloo},
  howpublished={arXiv preprint arXiv:2605.02210},
  year={2026}
}

@inproceedings{easynet,
  title = {{EasyNet}: 100 {Gbps} Network for {HLS}},
  author = {He, Zhenhao and Korolija, Dario and Alonso, Gustavo},
  booktitle = {International Conference on Field-Programmable Logic and Applications (FPL)},
  year = {2021}
}

@inproceedings{F4T,
  title={F4t: A fast and flexible fpga-based full-stack tcp acceleration framework},
  author={Boo, Junehyuk and Chung, Yujin and Baek, Eunjin and Na, Seongmin and Kim, Changsu and Kim, Jangwoo},
  booktitle={Proceedings of the 50th Annual International Symposium on Computer Architecture},
  pages={1--13},
  year={2023}
}

@inproceedings{lkl,
  title={{LKL}: The {Linux} kernel library},
  author={Purdila, Octavian and Grijincu, Lucian Adrian and Tapus, Nicolae},
  booktitle={9th RoEduNet IEEE International Conference},
  pages={328--333},
  year={2010}
}

@misc{fstack,
  title = {{F-Stack}: A high-performance user-space network framework based on {DPDK}},
  howpublished = {https://github.com/F-Stack/f-stack}
}

@inproceedings{eywa,
  title={Eywa: Automating Model-Based Testing using {LLMs}},
  author={Mondal, Rajdeep and Singha, Rathin and Millstein, Todd and Varghese, George and Beckett, Ryan and Kakarla, Siva Kesava Reddy},
  booktitle={23rd USENIX Symposium on Networked Systems Design and Implementation (NSDI 26)},
  year={2026}
}

\appendix

\section{Input-to-Event Mappings}
\label{app:event-mapping}

The mapping from concrete inputs to events is not one-to-one. A TCP program maps an acknowledgment to the same event whether it arrives in a pure ACK or piggybacked on a data segment, and a send to the same event whether the application calls \code{write} or \code{sendmsg}. The upper layers can even differ in kind: iWARP runs RDMA verbs over TCP \cite{rfc5040}, so sockets and verbs can map to one TCP program's events, and Falcon serves both InfiniBand verbs and NVMe over one transport \cite{falcon}. Which upper layer an operation comes in through is separate from the metadata the transport needs from it: the event declarations and the rest of the program do not change with the application API; only the parsers do.

The architecture file also bounds what parsers a stack can run: a hardware parser, e.g., is bounded in how deep into a packet it can look and how many fields it can extract. A protocol's packet formats and parsers must fit within these bounds.

Context lookup keys are also constructed in the parsers, and a context declaration states only the key's size (\sref{sec:state}). This is because how the key is constructed differs per event: the same TCP connection, e.g., is identified by an arriving segment's source address but by a send call's destination, so which fields form the key is meaningful only where the event is created.

\section{Factors in Managing Protocol State}
\label{app:state-factors}

When a stack decides how to store context instances and keep them consistent, it weighs the state's size, access pattern, and lifetime (\sref{sec:state}). The signposts below expand each into its fine-grained factors and mark each as a protocol decision the program states (\emph{protocol}), a workload or platform property the program does not carry (\emph{load}), or a split of the two. The goal is to show that \mtp program contain the protocol-related information needed for the stack to make its decisions.

\signpost{Size} The size of an instance and of its lookup key are determined by the protocol: the field list and the key size in the context declaration. The number of live instances, how many connections, RPCs, or flows are live, is determined by load. Variable-length fields, like lists of things maintained during a connection is a split of the two: the context declaration states whether a field size is only dynamically known (e.g., length of a list keeping track of pending connection waiting to get accepted), and the load determines the actual footprint.

\signpost{Access} The program states which instance an event accesses (the key construction in its parsers), whether each access reads or writes, and which fields are accessed together (both from the processing logic, \sref{sec:processing}). Key construction sets which events can reach an instance, but how many threads or cores access it concurrently depends on where the stack places event processing and on the load. How often events reach an instance, and which fields they touch most, are also determined by load.

\signpost{Lifetime} The program states which reactions create and remove an instance (\sref{sec:instructions}). How often instances are created and removed, and how long each lives, are determined by load.

\section{Code Optimizations}
\label{app:code-opts}

The same protocol processing behavior can be written in many semantically equivalent ways, and some may be more optimized for a particular stack than others.
For example, one common code optimization for TCP is to precompute the header for the next segment assuming common case (in-order, no unusual flags), and when the next segment arrives, instead of testing for each uncommon case in turn, do an equality check against the precomputed header to take the fast path if there is a match.
These are protocol-level decisions informed by the stack. \mtp{}'s intentional precision in describing processing logic allows a protocol expert to write the version tuned for a particular stack, and opens two avenues we leave to future work: a compiler that derives such versions automatically from a reference program, e.g., one written to match an RFC, and a checker that verifies an optimized version preserves the protocol's semantics.

\section{Example Segmentation Rules}
\label{app:seg-rules}

a TCP program can define a segmentation rule
\emph{[TCP\_bp:seq\_no, bp.hdr.seq\_no, prev.hdr.seq\_no + prev.payload\_len]}
(omitting \emph{last} for brevity), which instructs the target to segment the payload, set the sequence number of the first packet from the blueprint header, and increment it for each subsequent packet by the payload length of the previous packet.
Here, the \emph{bp} keyword refers to the blueprint instance, \emph{prev} to the previously generated packet, and \emph{.hdr} and \emph{.payload\_len} access the packet’s header and payload length.
As another example, in a \roce program, a segmentation rule
\emph{[RoCE\_bp::bth\_opcode, 0, 1, 2]}
can be used to set the opcode of the first, middle, and last packets generated from a single blueprint in response to an RDMA send event.

\section{More Details about Data Reassembly}
\label{app:data-factors}

\signpost{Optional address in \code{new\_rx\_ordered\_data}}
This can be used in application interfaces that already have a designated destination buffer before the data arrives, e.g., zero-copy interfaces like RDMA verbs. The application's buffer address will be a metadata in the event corresponding to the call, and the program passes it on when it creates the unit, which can lets a stack place each arriving segment at its final location with no intermediate copies.
This is different in a regular (non-zero-copy) socket interface. For example, in a TCP program, this optional address is not used at unit creation. When the application calls receive, the receive event will have the address as metadata and will use it in \code{rx\_flush\_and\_notify}.
Either way, an address states only where the data should eventually be available; whether the stack gets it there by directly by mapping addresses at unit creation time, or a copy at flush time remains the stack's decision.

\signpost{Semantics of the logical reassembly unit} The exact semantics of the reassembly state $(f, R)$ of \sref{sec:data-instrs} is as follows. \code{new\_rx\_ordered\_data} sets it to $(0, \emptyset)$. \code{add\_rx\_data\_seg(addr, len, uid, off)} sets it to $(f,\; \mathit{merge}(R, [\mathit{off}, \mathit{off}+\mathit{len})))$, where $\mathit{merge}$ adds the range to the set and coalesces ranges that overlap or meet, so $R$ stays the smallest set of disjoint ranges covering the added bytes. \code{rx\_flush\_and\_notify(uid, len, addr)} sets it to $(f+\mathit{len},\; R')$, where $R'$ is $R$ with all bytes below the new flush point removed. The constraints described in \sref{sec:data-instrs} are predicates over $f$ and $R$:  $\max(R) - f \leq W$ bounds how much memory is available for a buffer, $|R| \leq k$ bounds the extent of an out-of-order tracking, and a stack with no out-of-order tracking states that every added range must begin where the unit's in-order data ends.
This is the abstract semantics on top of \mtp{}'s language abstractions and, as discussed in \sref{sec:data-instrs}, does not prescribe how a stack should implement these instructions.

\section{More Details on \mtp Instructions and Constructs}
\label{app:other_instrs}

\signpost{Timer instructions}
Different stacks can have different ways of managing timers. For example, DPDK's timer library requires an explicit function call (\emph{rte\_timer\_manage}) to check the expiry of all timers local to a core and runs the callback functions for the expired ones. Other stacks like the Linux Kernel and XDP \cite{xdp-paper} do not require that explicit call, but XDP requires timers to be defined in eBPF maps. Some hardware stacks may have different ways of dealing with timers depending on their interval \cite{tonic}.
To interact with the stack for timer management, \mtp programs specify the timers that need to be instantiated for each connection in its per-flow context with a unique ID. At run-time, they use instructions such as \emph{timer\_start(tid, dur)}, \emph{timer\_restart(tid, dur)} and \emph{timer\_stop(tid)} to (re)start timer \emph{tid} with duration \emph{dur} or stop it.
The program will be notified of timer expiry through timer events with the timer id, and the corresponding registered event processing chains will be executed.

\signpost{``One-time'' scheduling instructions}
To specify its scheduling capabilities, the stack describes its available \emph{scheduling blocks} in its architecture file.
A scheduling block consists of a (configurable) number of queues and the scheduling policy mediating between them, and is expected to generate one stream of packets as outputs.
The stack will also specify if the scheduling blocks can be composed together by connecting the output of one block to one input queue of another.
Scheduling blocks can be instantiated and composed together by the protocol developer, and the result is communicated to the stack \emph{once} during deployment via \emph{register\_pkt\_sched(sched\_policy)}.

Specifically, a scheduling block of type \emph{S} can be instantiated with \emph{new\_sched\_block(S, queue\_cnt, params...)}, where \emph{queue\_cnt} is the number of queues in that scheduling block and \emph{params} are the specific scheduling parameters for that block, such as priorities or weights.
\emph{queue\_cnt} can be a concrete number, within the bounds specified by the stack, or \emph{PER\_FLOW} to ask the stack to create per-flow queues if supported\footnote{Scheduling blocks with per-connection flows cannot be downstream of other scheduling blocks in composition.}. In the latter case, the header fields used to distinguish individual flows (e.g., 4-tuple in TCP) should be provided as another parameter.

\signpost{Run-time scheduling instructions} Suppose a packet scheduler has been configured using the interface described above. At run-time, event processors can specify which queue the packets generated from a packet blueprint should go to in the \emph{pkt\_gen} instruction.
If supported by the stack, programs can also issue instructions to adjust scheduling parameters, such as the rate for rate limiters (e.g., for congestion control), or priorities at run-time using \emph{set\_queue\_rate(qid, rate)} and \emph{set\_queue\_prio(qid, prio)}, respectively.

\signpost{Event scheduling} Since the transport layer can receive events from concurrent event streams (e.g., the network, applications, and timers) and for multiple concurrent flows, there can be several events waiting to be processed at any time. The extent to which \mtp programmers can influence event scheduling depends on the stack. That is, similar to packet scheduling (\sref{sec:timer-instrs}), the stack can expose event scheduling blocks and allow developers to instantiate and compose them, and convey the result to the stack via \emph{register\_ev\_sched}.

\signpost{Notification} In addition to \emph{rx\_flush\_and\_notify} (\sref{sec:data-instrs}) and \emph{tx\_flush\_and\_notify} (\sref{sec:pktgen-instrs}), \mtp programs can use the \emph{notify} instructions to send other signals (e.g., completion signals in RDMA) to the application as well if needed.

\signpost{Sliding window data type, structs, and (sorted) lists} Sliding windows are commonly used in transport protocols to track and limit in-flight data for flow control and congestion control. As such, \mtp has a built-in sliding window data type to simplify specifying a protocol's operation at a high level. A sliding window in \mtp keeps track of a boolean flag for a sequence of numbers and has special functions to \emph{set} and \emph{unset} these flags for a range of numbers, find the \emph{first set/unset} flag, and \emph{slide} the window forward to a certain number.
\mtp also supports simple basic data structures, namely structs, lists, and sorted lists.
The stack is responsible for picking the most optimal data structure to implement them in its particular execution environment.

\section{\mtpdpdk and \mtpxdp details}
\label{app:stack-decisions}

\signpost{Packet generation}
eTran processes incoming packets in the eBPF hooks, keeps the send and receive buffers in a fast user-space data-path, and handles non-data upper-layer calls and timers in a slower user-space thread.
Consequently, packet generation, i.e., packet construction, payload access, and transmission, is split across several components depending on what triggers the packet and whether it carries data. We identified six such cases in eTran and the conditions behind them: (1) application-event-triggered data packet transmission, (2) data packet transmissions triggered by network events, e.g., duplicate ACKs, (3) control packets in response to data packets, (4) control packets in response to non-data packets, e.g., Homa busy replies, (5) pacing- or credit-based transmissions, and (6) transmissions triggered by control-path timer events.

Each case has its own optimized generation strategy, which eTran authors manually implemented for TCP and Homa. But nothing in the actual packet construction, payload access, or transmission needs to know what a TCP or Homa packet means. So, in \mtpxdp, we implemented each one once instead, driven only by a generic packet blueprint referencing header fields and payload bytes. Now, the conditions that determine which of the above packet generation strategies to pick, i.e., what triggered the issuing chain and whether payload is involved can be extracted by our compiler using standard static program analysis to generate the right packet-generation code for each \emph{pkt\_gen} call in the program.

In each iteration of its busy loop, after processing packets, upper-layer calls, and timeouts, mTCP iterates over flows with pending control packets, acks, and data, in that priority order, and generates packets accordingly.
\mtpdpdk{} keeps this batch-oriented strategy but generalizes it so that event processing marks flows that have issued \code{pkt\_gen} instructions by generic priority levels, as opposed organized around specific TCP packets, and the generation loop constructs packets from their blueprints.
Following eTran's optimized packet-generation path, for application-triggered data transmission, events originate and data is segmented in user-space, while the event processor that determines packet-generation parameters executes in XDP-EGRESS.
To bridge this boundary, \mtpxdp tags data segments with an application event ID and embeds event metadata in the first segment.
When event processors execute on that first segment in XDP-EGRESS, they generate and store the packet blueprint in an eBPF map; subsequent segments reuse this blueprint, introducing one additional lookup on the packet-generation path.

\signpost{State consistency}
In eTran, protocol processing is scattered across eBPF hooks and the two user-space paths (see above). So, each piece of transport state is either eBPF-only, user-space-only, or shared between the two sides, and eTran authors manually select the proper consistency mechanism for the shared ones: lock-free access for shared data that is only written by one side, and locks or thread-safe data structures otherwise. Moreover, eTran uses multiple cores, so for global state accessed when processing any flow's packets (e.g., Homa's grant scheduling variables), eTran synchronizes access using locks and atomic operations for simple ones, while for per-connection and per-RPC state, it shards and steers state across cores so there is no need for synchronization.

As with packet generation, we derive the appropriate synchronization mechanisms and the conditions that determine their use from eTran and use them in \mtpxdp. Once event processing is placed in eBPF hooks and user-space threads (above), there is sufficient information in the \mtp{} program, specifically, the granularities the state is organized into (\mtp{} contexts), which events access the same instance (key construction in \mtp{} parsers), and each field's readers and writers (static analysis of the processing logic),  for the \mtpxdp{} compiler to pick the right mechanism for each context instance and variable.

We derive \mtpdpdk{}'s synchronization strategy from mTCP in a similar process. Its sharing cases are fewer and simpler than eTran's. It runs an application thread and a stack thread on each core and takes a lock when the two share state, e.g., the send buffer; the little state shared across cores is protected with locks or atomic operations. \mtpdpdk{} keeps this setup, and the same program information as in \mtpxdp{} (\sref{sec:targets}) lets its compiler make the same selection, among these fewer cases, for each context variable.

\signpost{Send buffer adaptation to support packet generation optimization} mTCP's send buffer uses a consecutive array of bytes to implement a ring buffer. It keeps a pointer to the beginning of the array and another pointer to where the current head is. The array pointer does not change after initialization but the head pointer is moved forward when data is removed from the buffer (e.g., after an ack arrives).
Suppose the head is in the middle of the array, and we want to add data to the buffer. If the length of the data is less than the remaining spaces left until the end of the \emph{array}, mTCP copies the current data from the head to the beginning of the array, and then continues to add data after the new tail.
So, to reuse it in \mtpdpdk{}, we modify mTCP's send buffer to act as a ``true'' ring, where data is not moved, but head and tail pointers keep wrapping around.
This makes the packet-generation optimization in \sref{sec:soft-perf} safe: the referenced data does not move in \mtpdpdk's implementation of \code{tx\_data\_units} that represents send buffers.

\begin{figure}[t]
\centering
\setlength{\abovecaptionskip}{3pt}
\setlength{\belowcaptionskip}{-8pt}
\includegraphics[width=0.9\columnwidth]{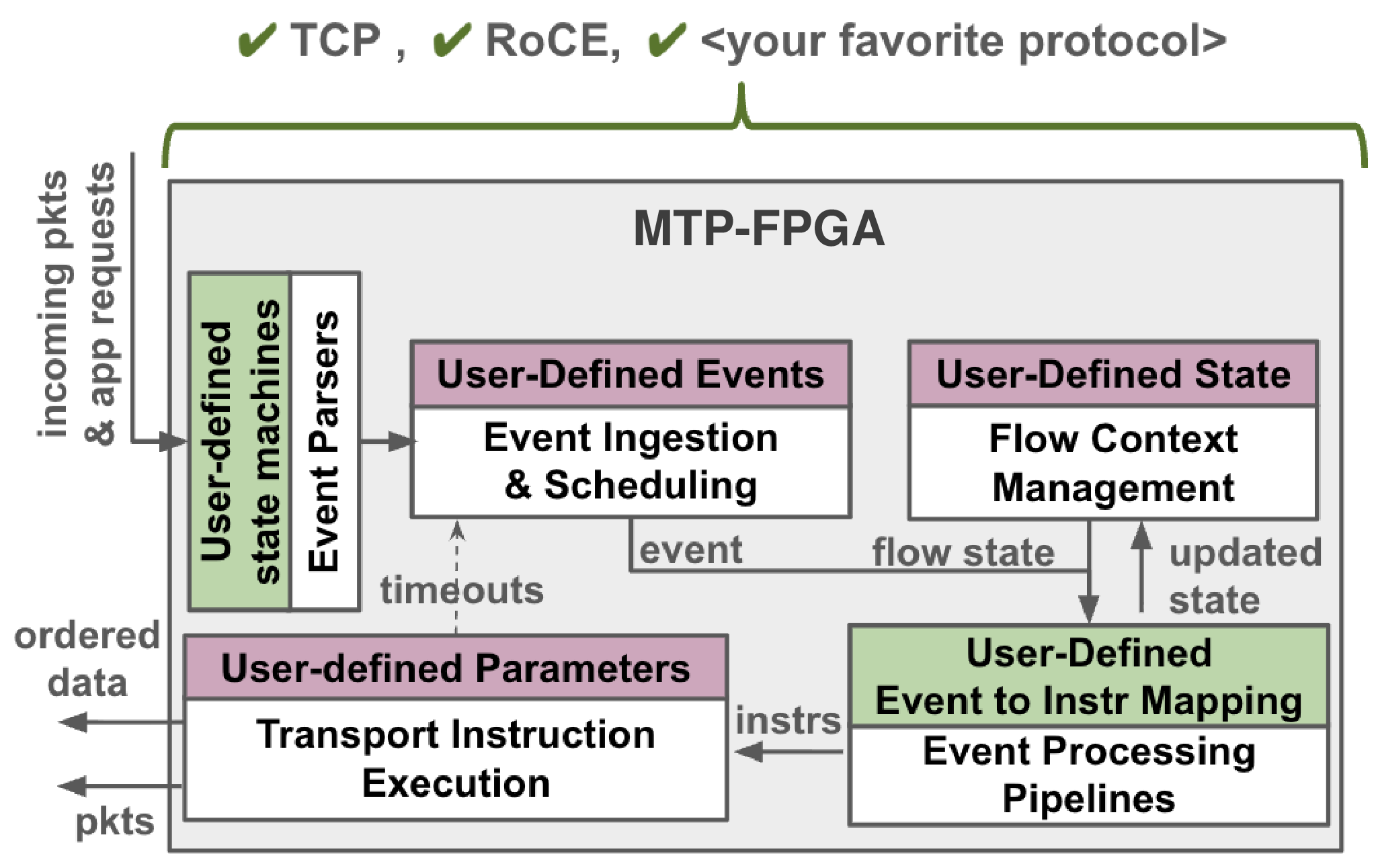}
\caption{\mtpfpga{}'s architecture supports protocols with radically different semantics (green: fully programmable, purple: protocol-agnostic and reconfigurable)}
\label{fig:pita-overview}
\end{figure}

\section{\mtpfpga{} Architecture and Event Processing}
\label{app:fpga-ple}

Figure~\ref{fig:pita-overview} shows \mtpfpga{}'s architecture: event ingestion and scheduling, and the packet generation, reassembly, and timer modules are implemented through protocol-agnostic mechanisms that operate on generic events and instructions, and the event processing pipeline is fully programmable to express protocol-specific logic within the \mtp abstraction.

We use TCP as a representative example of how \mtpfpga is programmed. Reprogramming it for \roce is similar.
To program \mtpfpga with the \mtp TCP program, the compiler should configure the event parsers in the event ingestion pipeline to extract the metadata required for TCP processing from incoming packets and socket send/receive requests.
The context table is configured with the per-connection state required for TCP processing, such as an integer tracking the first sent but unacked data segment or a bitmap tracking received segments.
The timer module is configured with TCP timers.
The compiler should create an C++ program (in HLS) for each event processing chain (i.e., each event). The HLS pipeline can then produce the pipelined Verilog implementation that fits into the rest of the modules and can together be synthesized to hardware.

As discussed above, programming \mtpfpga{} involves configuring the main datapath components and implementing the event processing logic using HLS.
We provided resource overhead results for the data path components in the main body of the paper. Table~\ref{tab:fpga-ple} summarizes the results for the HLS output for TCP* and \roce.
The two protocols differ significantly in the number of events, complexity of the event-processing logic, and pipeline depths, and how they parameterize transport instructions, reflecting their distinct semantics.
For example, \roce requires a larger set of events and more complex processing due to its operation-based model and richer packet semantics, while TCP relies on a smaller set of events but deeper processing pipelines.
Despite these differences, both protocols are implemented within the same protocol-agnostic datapath and incur modest hardware overhead.

\begin{figure}[t]
\centering
\setlength{\abovecaptionskip}{3pt}
\setlength{\belowcaptionskip}{-8pt}
\includegraphics[width=0.95\columnwidth]{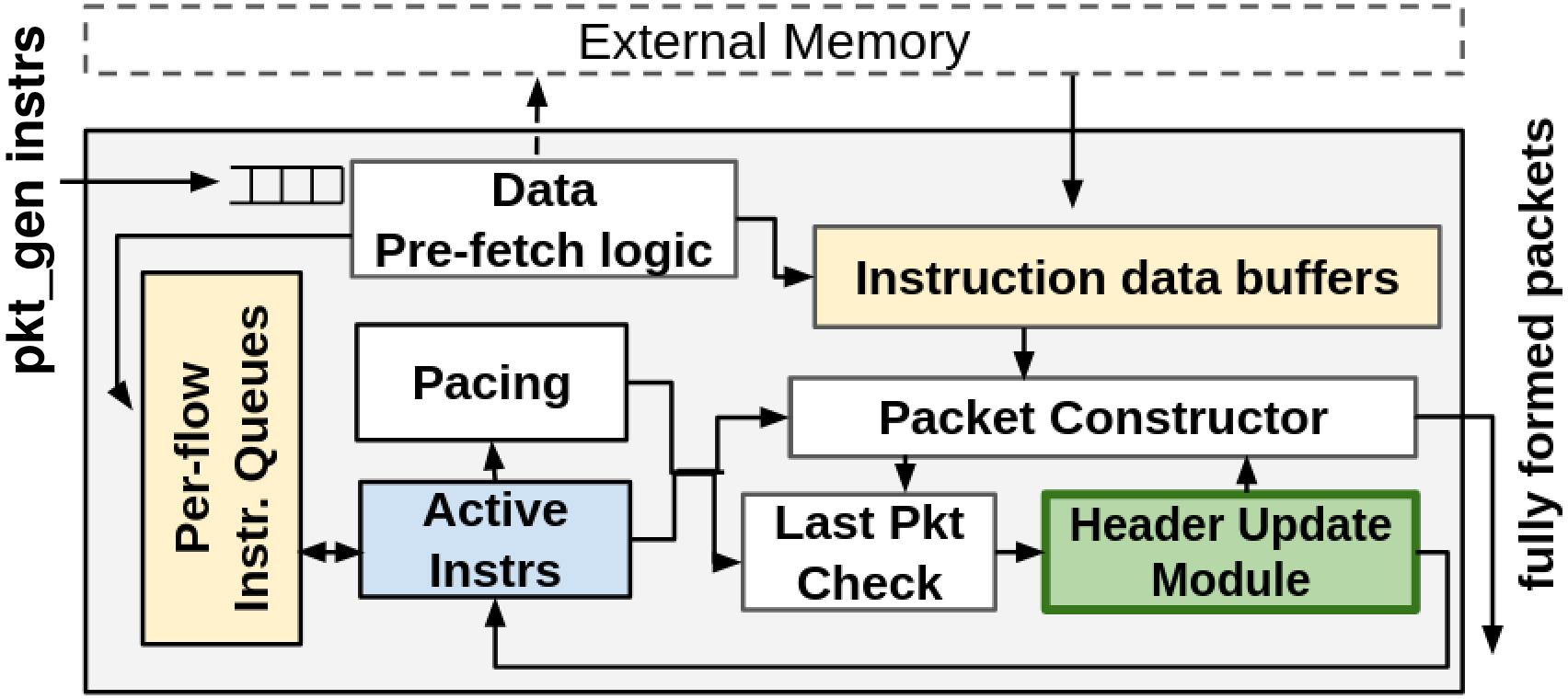}
\caption{\mtpfpga{}'s instruction-driven packet generation module. (green: reprogrammable, yellow: Dual-ported RAMs, blue: registers)}
\label{fig:pita-pktgen}
\end{figure}

\subsection{Packet generation module}

In \mtpfpga{}, packet generation makes no assumptions about header formats, data layout in memory, or how packets are derived from larger data units.
Instead, consistent with\mtp{}'s instruction abstraction (\sref{sec:instructions}), it is driven entirely by the packet generation instructions issued from the event-processing pipelines. Each instruction specifies all the information required to generate one or more packets, specifically the location and length of the data to be transmitted, the header to attach, the parameters needed to update the header across segments, and scheduling and pacing parameters such as per-flow rate or credit.

Figure~\ref{fig:pita-pktgen} shows a high-level overview of \mtpfpga{}'s packet generation module, which organizes execution around per-flow instruction queues, per-instruction data pre-fetch, and interleaved packet construction across flows.
Here, the definition of a flow is configurable to match the protocol's needs. For example, it may correspond to a TCP connection, a \roce{} queue pair, or an individual RPC message in Homa.
For each flow, the packet generator maintains a queue of instructions and tracks the progress of the currently active instruction.
Moreover, for each instruction, the packet generator maintains a data buffer that is continuously replenished with the payload data from the instruction's specified memory locations.
When an instruction is selected, the packet generator incrementally constructs packets from it according to the instruction parameters. If the instruction requires multiple packets, it remains active across multiple iterations until all data has been transmitted.
To sustain high throughput, the packet generator interleaves instruction execution across flows in a round-robin fashion.

\signpost{Continuous payload pre-fetch}
Data that goes into packet payloads is often stored in external memory.
\mtpfpga{} avoids stalling on external memory accesses by decoupling data retrieval from packet construction through continuous pre-fetching.
For each instruction, the packet generator maintains a dedicated payload buffer that is populated by a data-fetch engine. When an instruction arrives, data fetching begins immediately to retrieve payload data from the specified memory locations and store it in the buffer. The fetch engine continues to replenish the buffer in the background as packets are generated.
Specifically, further in the pipeline, packet construction incrementally consumes data from this buffer.
If the buffer's available data falls below a predefined threshold, the data fetch engine is notified for additional fetches, hiding memory access latency and allowing data retrieval and packet construction to proceed concurrently.

\signpost{Instruction arbitration}
The packet generator organizes instructions in per-flow ring buffers implemented using dual-ported RAMs and schedules packet transmission at flow granularity.
For each flow, \mtpfpga{} maintains the state of the currently active instruction along metadata such as execution progress and optional pacing parameters, specified in the instruction, in registers.
This information is used by a pacing module to schedule instructions for packet construction.
To avoid head-of-line blocking, an instruction can be preempted after generating a configurable amount of data or packets to return from packet construction, update its execution state, and be rescheduled by the pacing module.

\signpost{Customizable segmentation and packet construction}
The packet constructor receives the parameters of the selected instruction from the scheduler, consumes payload data from the per-instruction buffer, and generates packets according to the specified segmentation parameters.
For each packet, it reads one segment of pre-fetched data (with segment size defined by the instruction), attaches the specified header, and transmits the resulting packet.
If the instruction parameters describe more data than fits in a single packet, the constructor iteratively produces multiple packets, consuming buffered data and triggering additional data fetches as needed to maintain continuous execution.

A key component of this process is a configurable header-update module that determines how header fields evolve across segments.
Rather than assuming a fixed update rule, the module can be configured to apply protocol-specific rules that take in the current header, bytes transmitted so far from this instruction, and instruction parameters, and produce the next header.
For example, in TCP, this component can be configured to add the instruction segment size to the sequence number of the current header to derive the sequence number for the next packet. For \roce, it can be configured to set opcodes depending on whether the packet is the first, the middle, or the last one in an operation.
\begin{table}[t]
\small
\centering
\setlength{\abovecaptionskip}{3pt}
\setlength{\belowcaptionskip}{-8pt}
\begin{tabular}{@{}lcc@{}}
\toprule
 & \textbf{TCP*} & \textbf{\roce} \\
 & \textbf{(w/ AIMD)} & \textbf{(w/ DCQCN)} \\
\midrule
User-defined events & 4 & 17 \\
HLS LoC & 612 & 1434 \\
Pipeline depth (max) & 12 & 6 \\
Pipeline depth (avg) & 4.75 & 2.35 \\
FF usage & 13,384 ($<$1\%) & 132,731 (3.8\%) \\
LUT usage & 7,346 ($<$1\%) & 51,972 (3\%) \\
\bottomrule
\end{tabular}
\caption{TCP* and \roce{} event-processing logic on \mtpfpga{}. Percentages are of an AMD Alveo U250's resources.}
\label{tab:fpga-ple}
\end{table}

\subsection{Data reassembly module}
Consistent with \mtp{}'s instruction abstraction (\sref{sec:instructions}), \mtpfpga{} does not make any assumptions about ordering or completion. Instead, the program's event processors explicitly specify these semantics through reassembly instructions.

When data packets pass through the programmable parser, their payloads are temporarily stored in a ring buffer in memory in arrival order.
For each packet, the address of its corresponding payload in that memory is provided to the event processor as part of the event metadata.
Based on protocol-specific logic, the event processor determines the correct placement of each segment and issues an \emph{add\_rx\_data\_seg} instruction with the address of the segment in the temporary memory and the correct offset as parameters.
The reassembly module then fetches the payload from the temporary memory and inserts it at the specified offset in a per-flow reassembly buffer.
The reassembly module also supports the \emph{rx\_flush\_and\_notify} instruction, which makes a contiguous portion of the buffer available to the application.
In both cases, the module performs only the operations specified in the instruction, without tracking segment state or inferring ordering, leaving all protocol semantics to the event processors.

\signpost{Handling segment alignment}
Executing the \emph{add\_rx\_data\_seg} instruction involves fetching the segment from temporary payload memory and inserting it into the reassembly buffer in a dual-ported RAM at the specified byte offset.
Reassembly buffers are organized in memory-addressable fixed-size chunks.
Because \mtpfpga{} does not assume any alignment guarantees from instructions, the reassembly module must support arbitrary byte offsets.
To do that, it aligns incoming data with chunk boundaries using a shift pipeline and performing read-modify-write for the first and last partial chunks when needed.
This process is largely pipelined, where inserting a segment spanning $N$ chunks takes $N+1$ cycles. The extra cycle arises from misalignment and boundary reads, needed to support arbitrary offsets.
Larger chunk sizes require supporting a wider range of shifts; we use 64B chunks, matching the minimum packet size.

For example, consider inserting a 256B segment at offset 71 into a buffer with 64B chunks.
The shift amount is given by the offset modulo the chunk size, i.e., $71 \bmod 64 = 7$.
The segment spans four chunks, which are passed through a shift pipeline that applies shifts based on the set bits in the bit representation of the shift amount (here, shifting by 1, 2, and 4 bytes).
Starting from the second chunk, the pipeline concatenates the previous and current chunks, shifts them together, and selects the first 64 bytes of the result. This ensures that any leftover data from shifting the previous chunk is correctly accounted for.
For the first and potentially last partial chunks, existing buffer contents must be preserved. In this example, the first 7 bytes of the first chunk and the last 57 bytes of the final chunk are read from memory, merged with the shifted data, and written back.

\signpost{Flush and notify}
Once an event processor determines that a contiguous region of data of size $X$ is ready, it issues an instruction to transfer $X$ more bytes from the reassembly buffer to an application-provided memory address.
To support this, the reassembly module maintains a per-flow read pointer into the buffer and, upon receiving an  \emph{rx\_flush\_and\_notify} instruction, outputs the required number of chunks starting from this pointer and advances it so that it always points to the first chunk with data not yet exposed to the application.
This allows the reassembly module to provide the required data for transfer to the application solely based on instruction parameters and without interpreting protocol-specific completion conditions.

\subsection{More explanation of microbenchmark results} In the microbenchmarks of \sref{sec:hw-perf},
for minimum-sized packets (64B), the current packet generator prototype does not fully sustain line rate under back-to-back single-packet instructions.
This is because transmission of a 64B packet completes in one cycle, requiring the packet constructor to wait for the next instruction for updated header and segmentation parameters. For larger packets, this latency is hidden by packet transmission time.
This gap does not reflect a fundamental limitation and can be addressed using standard techniques such as instruction lookahead or prefetching.
Moreover, this case arises only when flows issue back-to-back instructions, each corresponding to a single minimum-sized packet.
Such patterns are uncommon in well-formed transport protocols, which, instead of issuing back-to-back individual small packets, batch data into multiple full-sized packets.

For smaller segments (128B and below), the current prototype for data reassembly module does not fully sustain line rate under back-to-back instructions. This is due to an additional cycle required when segment offsets are not aligned with the 64B chunk boundaries of the reassembly buffer, which introduces occasional read-modify-write overhead.
In practice, workloads rarely consist solely of back-to-back small segments, and mixed traffic allows this overhead to be amortized, enabling the reassembly module to sustain line-rate operation.

\end{document}